\documentclass[twocolumn,aps,prb,amsmath,superscriptaddress,notitlepage,longbibliography]{revtex4-1}
\usepackage{amssymb}
\usepackage{mathrsfs}
\usepackage{graphicx}
\usepackage{grffile}
\usepackage{float}
\usepackage[caption=false]{subfig}
\usepackage{scalerel}
\usepackage[pdftex,colorlinks=red]{hyperref}
\usepackage{amscd}
\usepackage{graphicx}
\usepackage{epstopdf}
\usepackage[normalem]{ulem}
\usepackage{verbatim}
\usepackage{xcolor}
\usepackage{bm}
\usepackage{dcolumn}
\usepackage[utf8x]{inputenc}
\definecolor{Gray}{gray}{0.9}

\def\blue{\textcolor{blue}}

\begin{document}
	
	\def\qv{\vec{q}}
	\def\blue{\textcolor{blue}}
	\def\magenta{\textcolor{magenta}}
	\def\apricot{\textcolor{Apricot}}

	\definecolor{ora}{rgb}{1,0.45,0.2}
	\def\LH{\textcolor{black}}

	\newcommand{\norm}[1]{\left\lVert#1\right\rVert}
	\newcommand{\ad}[1]{\text{ad}_{S_{#1}(t)}}
	
	\title{Real non-Hermitian energy spectra without any symmetry}
	\author{Boxue Zhang}
	\email{These authors contributed equally to the manuscript.}
	\affiliation{School of Physics, Sun Yat-sen University, Guangzhou 510275, China}
	\author{Qingya Li}
	\email{These authors contributed equally to the manuscript.}
	\affiliation{Department of Physics, National University of Singapore, Singapore 117542}
	\author{Xiao Zhang}
	\affiliation{School of Physics, Sun Yat-sen University, Guangzhou 510275, China}
	\author{Ching Hua Lee}
	\email{phylch@nus.edu.sg}
	\affiliation{Department of Physics, National University of Singapore, Singapore 117542}
	

	\date{\today}
	\begin{abstract}
		Non-Hermitian models with real eigenenergies are highly desirable for their stability. Yet, most of the currently known ones are constrained by symmetries such as PT-symmetry, which is incompatible with realizing some of the most exotic non-Hermitian phenomena. In this work, we investigate how the non-Hermitian skin effect provides an alternative route towards enforcing real spectra and system stability. We showcase, for different classes of energy dispersions, various ansatz models that possess large parameter space regions with real spectra, despite not having any obvious symmetry. These minimal local models can be quickly implemented in non-reciprocal experimental setups such as electrical circuits with operational amplifiers.
	\end{abstract}
	
	\maketitle

	\section{Introduction}
	
	Non-Hermitian systems has recently inspired intense research efforts for their unconventional mathematical properties and physical robustness, such as enlarged topological symmetries~\cite{Kawabata2018nonHclass,Liu2019nonHclass,bergholtz2021exceptional,sayyad2021entanglement,wojcik2021eigenvalue,shiozaki2021symmetry,borgnia2020nonH,park2022nodal}, exceptional point sensing~\cite{chen2017exceptional,wiersig2014enhancing,zhang2019quantumnoise,Zeng2019enhancedsensitivity,Budich2020NonHermitian,sahoo2022tailoring,hodaei2017enhanced}, quantized classical responses~\cite{li2021quantized}, modified bulk-boundary correspondences~\cite{yao2018edge,yang2020non,yokomizo2019non,helbig2020generalized,Yao2018nonH2D,zhang2019correspondence,guo2021analysis,bartlett2021illuminating,zhang2022bulkbulk,cao2021non,Song2019BBC,kunst2018biorthogonal,koch2020bulk,imura2020generalized,qin2022non,jiang2022filling}, unconventional entanglement entropy scaling~\cite{li2021critical,lee2022exceptional,sayyad2021entanglement,chen2022characterizing,chang2020entanglement,zhou2020renormalization,yokomizo2021scaling,pan2021point}, enhanced Rabi oscillations~\cite{lee2020ultrafast,cao2021non,zhou2021dual}, effective non-Hermitian curved space~\cite{lv2021curving,marcus2022pt}. Yet, many of these exciting phenomena are often difficult to probe experimentally due to their intrinsically unstable nature from complex eigenenergies. While real eigenenergies can be symmetry-enforced i.e. through PT-symmetry~\cite{Bender1998nonH,el2018non,stegmaier2021topological,fring2022introduction,xiao2021observation,Schomerus2013topo,Weimann2016topo,zdemir2019ParitytimeSA}, doing so is incompatible with realizing many of the most exotic non-Hermitian phenomena~\cite{ghatak2020observation,helbig2020generalized,xiao2020non,Bouganne2020anomalous,song2019breakup,hou2020topo,longhi2018paritytime,ningyuan2015time,scheibner2020non,zeng2021real,long2021realeigenvalued,kawabata2020real,aharonov1996adiabatic,nirala2019measuring,Li2021protecting}.
	
	In this work, we carefully investigate how the non-Hermitian skin effect (NHSE)~\cite{Martinez2018nonH,okuma2020topological,longhi2021non,peng2022manipulating,liang2022observation,zhu2020photonic,guo2021exact} can also enforce real non-Hermitian spectra, even for lattices whose couplings and momentum-space descriptions do not admit any obvious symmetry. The NHSE has been heavily associated with modified bulk-boundary correspondences~\cite{yao2018edge,pan2021point} and, in our context, implies that a system can robustly possess real spectrum in the presence of a boundary, even though its bulk is unstable with complex eigenenergies. Physically, this is because the directed amplification in a NHSE lattice can be stabilized by the interfering wavepackets from a boundary or spatial inhomogeneity, a mechanism that is unrelated to conventional symmetry protection.

	We elucidate this route towards real eigenspectra in terms of the inverse skin depth~\cite{yao2018edge,lee2020unraveling}$\kappa$, which is an additional degree of freedom that mathematically takes the role of imaginary momentum. It physically controls the accumulation and interference of skin states at a boundary, and mathematically replaces the Bloch description of lattice by an effective \emph{surrogate}~\cite{lee2020unraveling} model that can look completely different. Specifically, we shall show that in a bounded lattice, the reality of the spectrum is only destroyed by a spontaneous symmetry breaking process that can occur much later than the explicit symmetry breaking at the Bloch level.

	\section{OBC vs PBC spectra}
	
	We first review and distinguish the approaches for computing the eigenenergy spectrum under open vs. periodic boundary conditions (OBCs vs PBCs). Given a generic non-Hermitian Hamiltonian $H(k)$, PBC eigenenergies $\tilde E$ and OBC eigenenergies $\bar E$ are obtained very differently. To find the set of $E\in\tilde E$, we simply solve for eigenenergies $E$ such that characteristic polynomial $P(E,k) = \text{Det}[H(k)-E\,\mathbb{I}]=0$, where $k\in [0,2\pi)$. However, under OBCs, translation invariance is broken, and in general the spectrum is \emph{not} indexed by real momentum $k$. Instead, the OBC spectrum $E\in \bar E$ is given by eigenenergies $E$ that solve
	\begin{equation}
		P(p)=P(E,k+i\kappa) = \text{Det}[H(k+i\kappa)-E\,\mathbb{I}]=0
	\end{equation}
	and are \emph{degenerate in both} $E$ and $\kappa$. Here $\kappa$ is the imaginary part of the complexified momentum $p=k+i\kappa$ that also represents the inverse decay length (skin depth) of eigenstates viz. $e^{ipx}\sim e^{-\kappa x}$ i.e. taking the role of a length scale~\cite{qi2013exact,gu2016holographic} not present in Hermitian systems. This $\kappa$ degeneracy is required because OBC skin eigenstates have exponential spatial profiles, and we need a superposition of two of them with identical $E$ and $\kappa$ to satisfy OBCs at two arbitrarily separated boundaries~\footnote{As we interpolate between OBCs and PBCs, we observe a peculiar scaling behavior of the corresponding effective $\kappa$~\cite{li2021impurity}.}. In general, we write $\kappa=\kappa(k)$ to emphasize its $k$-dependence, and $p=k+i\kappa(k)$ is known as the generalized Brillouin zone (GBZ)~\cite{yao2018edge, yang2020non, yokomizo2019non, Yao2018nonH2D, zhang2019correspondence, guo2021analysis,guo2021non,yang2020non}. $H(p)=H(k+i\kappa(k))$ is also referred to as the surrogate Hamiltonian, which is used instead of the original Bloch Hamiltonian $H(k)$ in computing topological invariants~\footnote{However, the topological eigenenergies themselves fall outside of the purview of our prescription, because they are isolated solutions that are not adiabatically connected to any Bloch solution.}~\cite{ghatak2019new,song2019realspace,liu2021real} and spectral properties under OBCs.
	
	Note that the above prescriptions for the OBC and PBC eigenenergies cocide in the case of Hermitian lattices, since as $p$ cycles through real values $[0,2\pi)$, every value of $E$ lies on the real line and will be visited at least twice, both with $\kappa =\text{Im}(p)=0$. 
	
	\subsection{Minimal model with different OBC vs PBC spectra}
	
\begin{figure}
		\includegraphics[width=\linewidth]{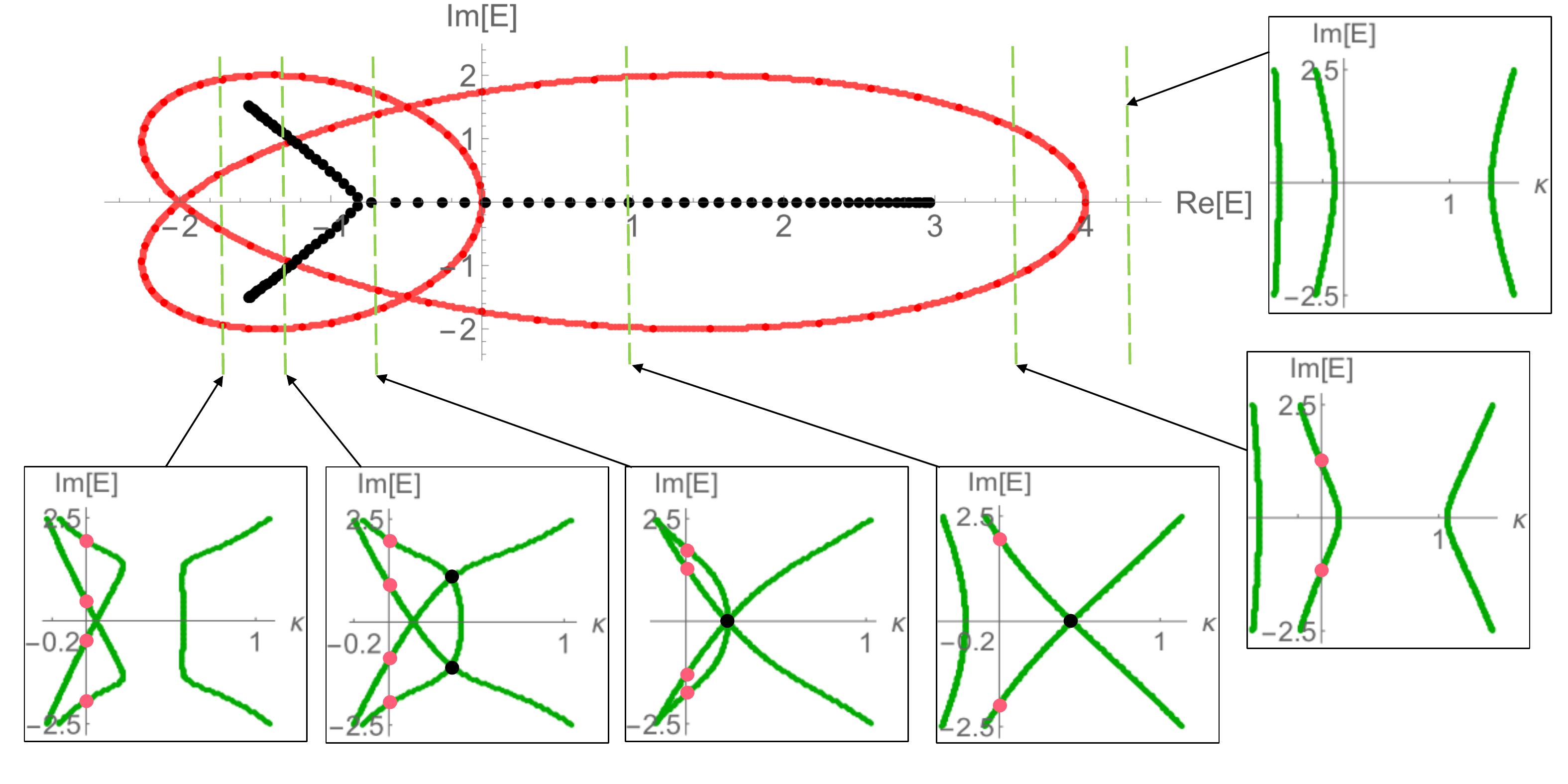}\\
		\caption{Real OBC vs. real PBC spectra in terms of the symmetry of $\kappa(E)$ solutions. Shown are the OBC (black) and PBC (red) spectra of the model $H_\text{min}$ with dispersion $E(z)=z+z^{-1}+2z^2$; the inset plots shows $\text{Im}(E(z))$ as a function of $\kappa=-\log|z|$ (green $\kappa$ curves) at various fixed $\text{Re}(E(z))$ slices. PBC eigenenergies (red dots) correspond to $(\text{Re}(E(z)),\text{Im}(E(z)))$ consistent with $\kappa=0$, and OBC eigenenergies (black dots) correspond degeneracies in both $\kappa$ and $E(z)$ (green curve intersections). In particular, we have real OBC eigenenergies when the green curves intersect at $\text{Im}(E)=0$, which often occurs even when the PBC eigenenergies are non-real. }
		\label{fig:1} 
	\end{figure}

	As a concrete demonstration, we consider a minimal model with $H_\text{min}(z)=z+\frac1{z}+Az^2$, where $z=e^{ik}$. In real space, it contains two symmetric nearest neighbor (NN) hoppings and another uncompensated next-nearest-neighbor (NNN) hopping: $H_\text{min}=\sum_x |x+1\rangle\langle x| + |x\rangle\langle x+1|+A|x+2\rangle\langle x|$. Clearly, its PBC spectrum is given by $\tilde E = 2\cos k +Ae^{2ik}$, and is entirely complex unless $A=0$, as plotted as the thick red curve in Fig.~\ref{fig:1} for $A=2$. However, large segments of its OBC spectrum lie on the real line, as shown by the black dots. 
	
	Below, we explain how one can visually derive the OBC and PBC spectra. We turn to the plots of $\kappa$ solutions vs. $\text{Im}(E)$, for fixed $\text{Re}(E)$ slices. Going from large to small $\text{Re}(E)$, we find that PBC eigenenergies (red) first appear when we pass $\text{Re}(E)=4$, followed by OBC eigenenergies (black) after $\text{Re}(E)=3$. In the green $\text{Im}(E)$ vs. $\kappa$ plots, it is apparent that PBC eigenenergies appear when the green $\kappa$ solution curves cross $\kappa=0$, while OBC eigenenergies only appear when the $\kappa$ curves intersect. This is exactly what was prescribed earlier - $\kappa=0$ gives the PBC spectrum, while $\kappa$ degeneracies give the OBC spectrum. As $\text{Re}(E)$ decreases further, additional PBC winding loops appear, and they correspond to additional $0$ crossings of $\kappa$ from the green curve that is emerging from small $\kappa$. At $\text{Re}(E)\approx -0.79$, that green curve goes over the original black $\kappa$ intersection, thereby splitting it into two black intersections. That corresponds to  the two black OBC branches away from the real $E$ line. Finally, when these two $\kappa$ intersections gap out at sufficiently negative $\text{Re}(E)$, the OBC eigenenergies disappear.  
	
	All in all, we see that real OBC eigenenergies correspond to intersections of the $\kappa$ curves at $\text{Im}(E)=0$, which can exist even if PBC eigenenergies are already complex i.e. if the $\kappa$ curves cross $\kappa=0$ at $\text{Im}(E)\neq 0$. As such, the breakdown of the reality of the spectra can be understood as the breaking of the symmetry of the $\kappa$ curves. While a non-real PBC spectrum only requires the $\kappa$ curves to have asymmetric zero crossings, which are almost guaranteed in a system without PT symmetry, a non-real OBC spectrum requires that symmetry to be \emph{spontaneously broken} i.e. broken at the level of ``extrema'' corresponding to the intersection points. Tellingly, it is often much harder to have asymmetric $\kappa$ intersections rather than $\kappa$ zero crossings, and that explains the relative robustness of real OBC spectra compared to real PBC spectra.

	\section{Parameter spaces for real eigenenergies}
	
	We next present the parameter space for real OBC spectra for several paradigmatic models. It has to be emphasized that almost all eigenenergies (under both OBCs and PBCs) fundamentally depends on the form of the dispersion $P(E,p)$, and only indirectly on the form of the Hamiltonian. The exceptions are the eigenenergies of isolated topological modes, which are protected by bulk eigenstate topology, but we will not focus on them here. 
	
	An important simplifying symmetry for OBC (but not PBC) spectra is its invariance under constant translations of $\kappa$ i.e. $H(p)$ and $H(p+i\kappa_0)$ always possess identical OBC spectra for any fixed $\kappa_0$. This is because the OBC spectra is determined by $\kappa$ crossings, which are unaffected by overall translations. As such, each model belongs to an equivalence class of models related by hopping rescalings, all possessing the same OBC spectrum. As an illustration, $H_\text{min}(z)=z+\frac1{z}+Az^2$ has identical OBC spectrum as $H'_\text{min}(z)=e^{-\kappa_0}z+\frac{e^{\kappa_0}}{z}+Ae^{-2\kappa_0}z^2$, for which the $A=0$ case reduces to the famed Hatano-Nelson model with unequal NN couplings~\cite{koch2020bulk,gong2018topological,zhang2022symmetry}.

	\subsection{Separable energy dispersions $P(E,p)$}
	We first discuss separable dispersions, namely those with $P(E,p)=F(E) + G(p)$, where $F(E)$ is a function of $E$ and $G(p)$ is a Laurent polynomial of $z=e^{ip}$. As long as $G(p)$ gives an OBC spectral curve that does not contain branches, we can in principle achieve a real spectrum by modifying the model such that $F(E)$ conformally~\cite{tai2022zoology} maps the curve onto the real line. For this reason, the non-Hermitian SSH model and its variants can all possess real spectra~\cite{yao2018edge,Lee2019anatomy,tai2022zoology}.
	
	\subsubsection{Single-component Hamiltonians}
	We start with the single-component Hamiltonians, whose characteristic polynomials are simply given by $P(E,p)=H(p)-E$. As discussed above, cases with only two NN hoppings are trivial, since they are always reducible to the equivalence class of $H(z)=z+z^{-1}$. As such, the minimal nontrivial case is the 3-hoppings model $H_\text{min}(z)$, which we just examined. There are two ways to generalize to the next level of sophistication through a fourth hopping term, namely
	\begin{equation}
		H_\text{1-band}^1(z)=z+\frac1{z}+Az^2+Bz^3  ,
	\end{equation}
	\begin{equation}
		H_\text{1-band}^2(z)=z+\frac1{z}+Az^2+\frac1{z^2}{B} 
	\end{equation}
	where $z=e^{ip}$. These two models capture all the possibilities for Hamiltonians with hoppings spanning four sites, up to reflection and translation symmmetry. Note that for single-component models, the onsite term is just a trivial constant. Also, all meaningful models must possess both left and right hoppings, since otherwise the OBC spectrum will collapse onto a single point~\cite{longhi2020non,martinez2018non}.
	
	\begin{figure}
		\subfloat[]{\includegraphics[width=.4\linewidth]{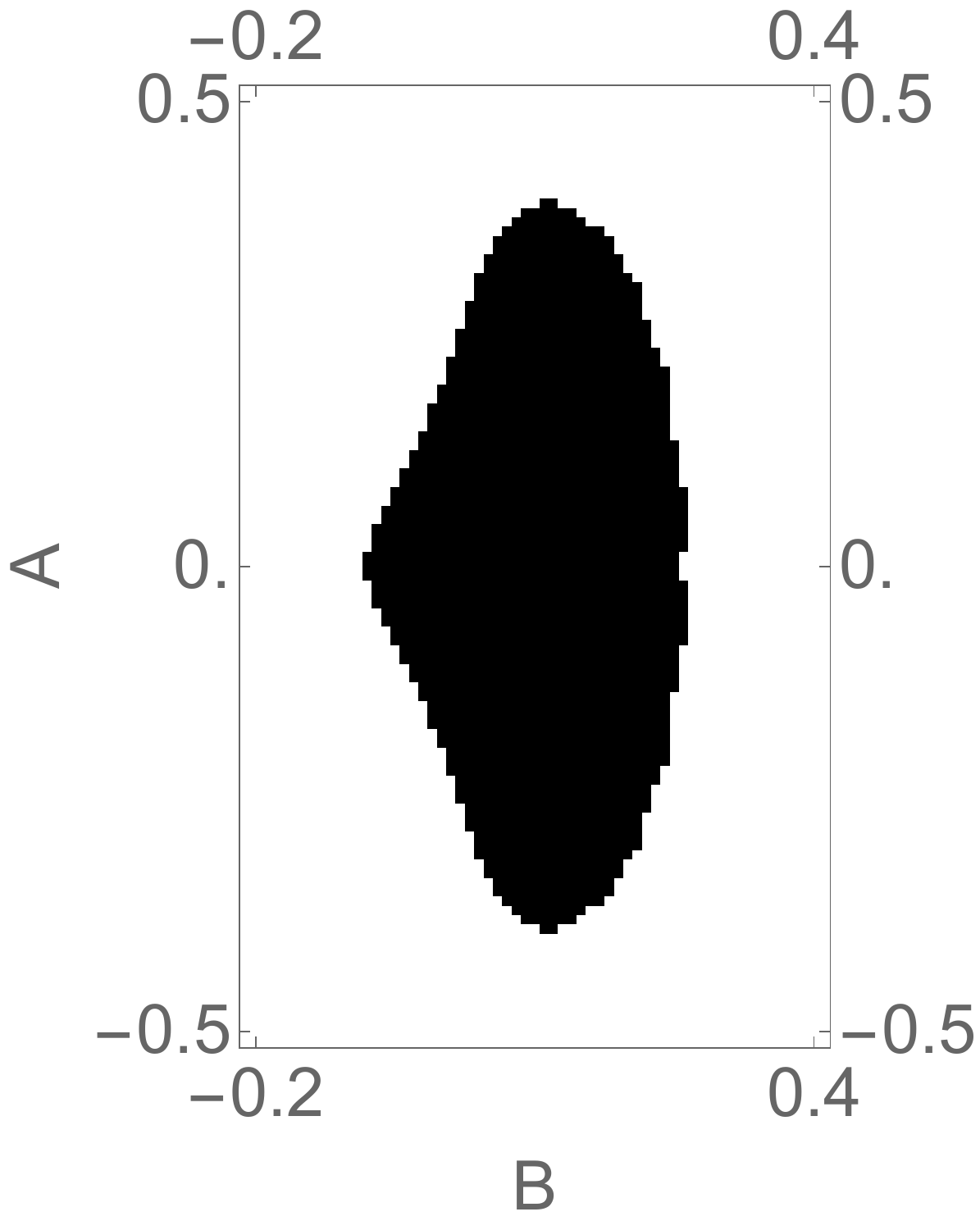}}
		\subfloat[]{\includegraphics[width=.49\linewidth]{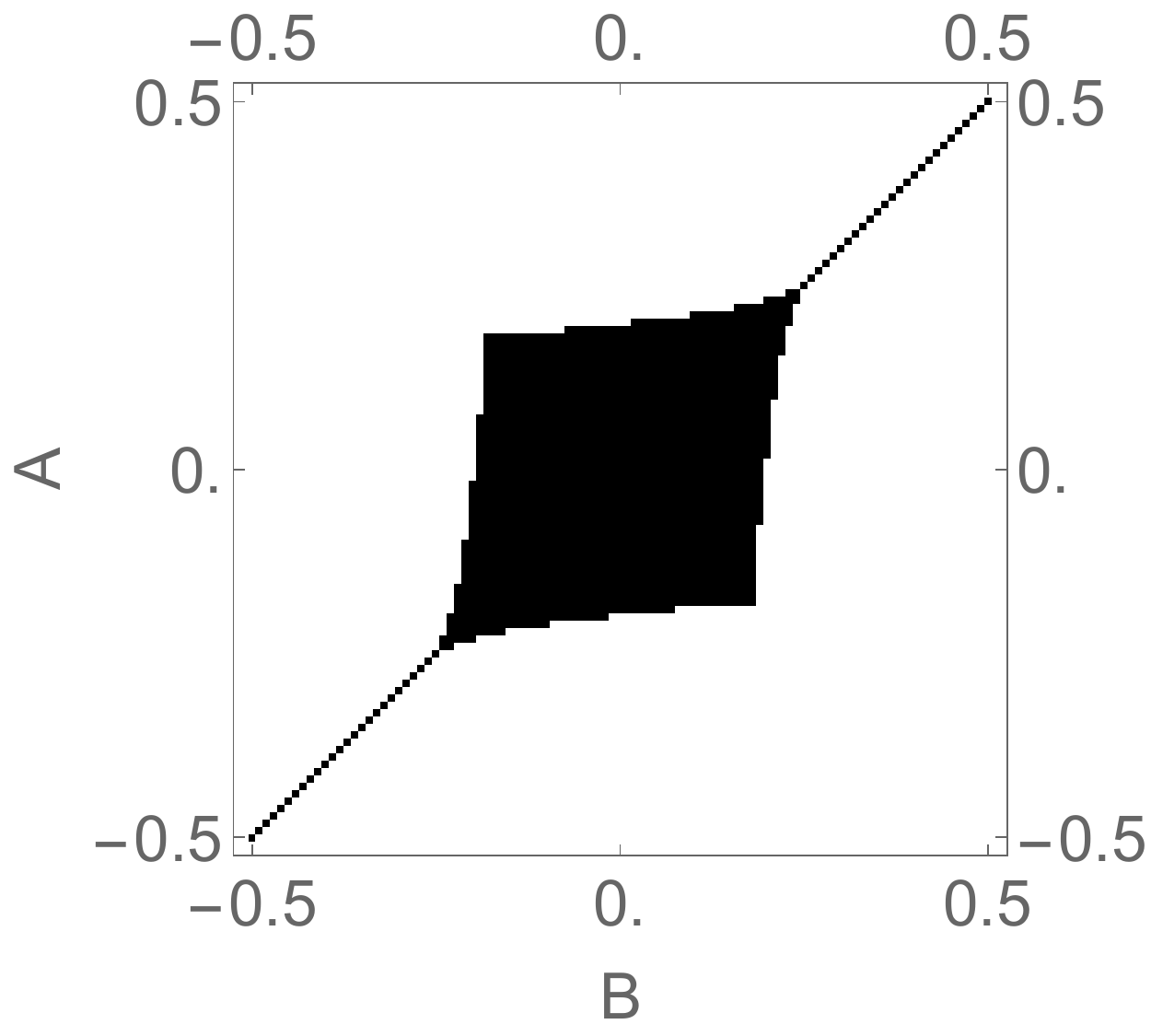}}
		\caption{
			Regions of the parameter space with real OBC spectra (black) for the following 2-component models: (a) $H_\text{1-band}^1$ and (b) $H_\text{1-band}^2$. In (b), models with $A=B$ also trivially result in real spectra. The numerical threshold is  $\text{Max}|\text{Im}(E)|<\epsilon=10^{-6}$.}
		\label{fig:2} 
	\end{figure}
	
	As we can see from Figs.~\ref{fig:2}a and b, there exists a rather large (black) region in the $(A,B)$ parameter space where the spectra still remain real, despite $A$ and $B$ manifestly breaking any possible symmetry. Indeed, the $Az^2+Bz^3$ term of $H_\text{1-band}^1(z)$ gives rise to robustly complex eigenenergies under PBCs, even though it can still give a real spectrum for $A$ as large as $0.35$ (Fig.~\ref{fig:2}a). For $H_\text{1-band}^2(z)$ with dispersion $2\cos p +Ae^{2ip}+Be^{-2ip}$, the OBC spectrum is trivially real for $A=B$, but still remains real for a large parameter region away from that (Fig.~\ref{fig:2}b). Physically, that is so because interference from waves reflected off a boundary are sufficient in preventing a wavepacket from being amplified indefinitely. 
	
	\subsubsection{Two-component Hamiltonians}
	In 2-band models, we have 
	\begin{equation}
		P_2(E,p)=E^2 - [\text{Tr}H_\text{2}(p)]E + \text{Det}H_\text{2}(p),
		\label{P2Ep}
	\end{equation}
	such that separable dispersions correspond to Hamiltonians with $p$-independent traces, which can occur when the diagonal terms are either zero or constant. Physically, this corresponds to the absence of homogeneous same-sublattice net hoppings across different unit cells.
	
	\begin{figure*}
		\subfloat[]{\raisebox{-0.5\height}{\includegraphics[width=.24\linewidth]{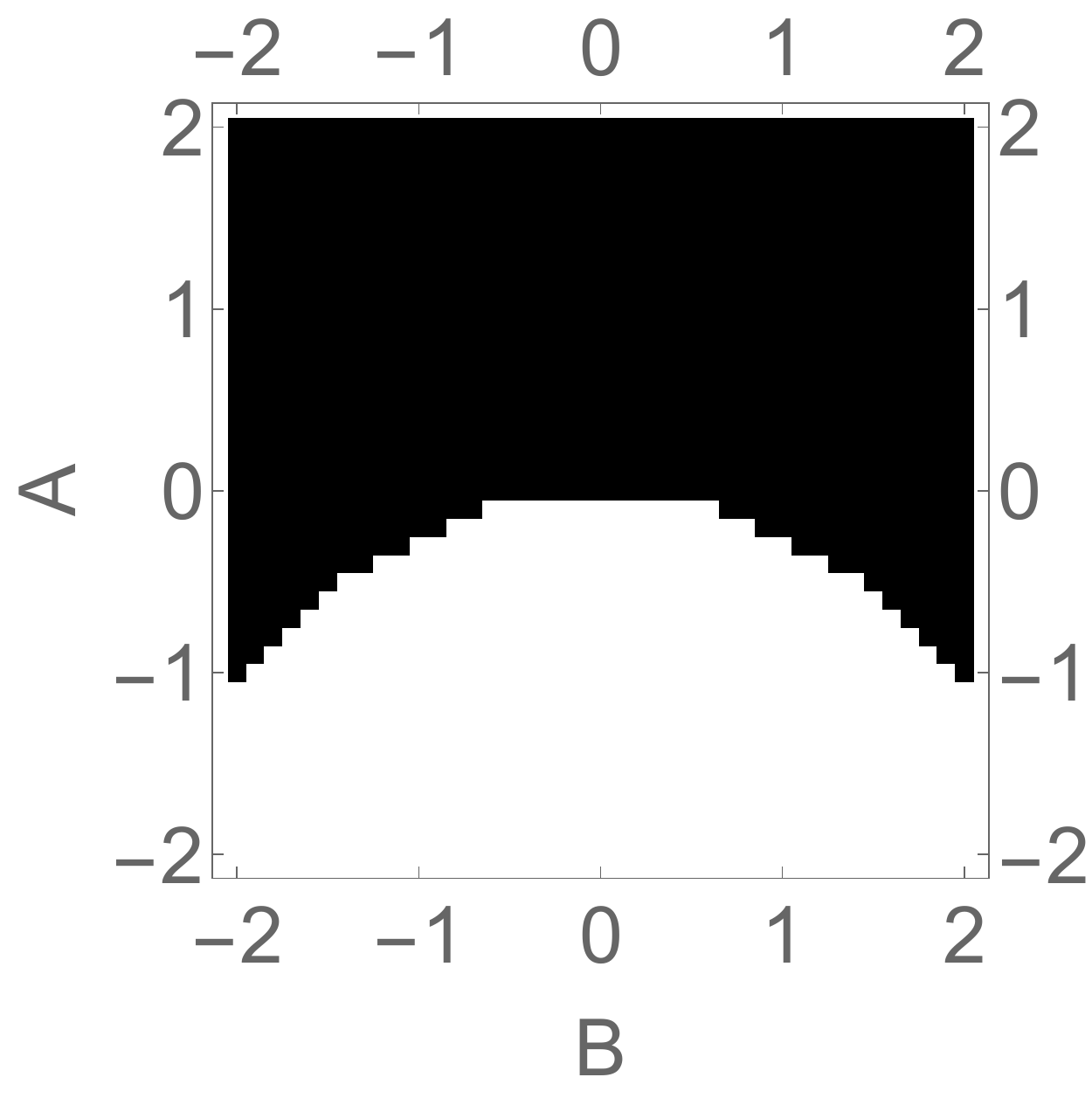}}}
		\subfloat[]{\raisebox{-0.5\height}{\includegraphics[width=.24\linewidth]{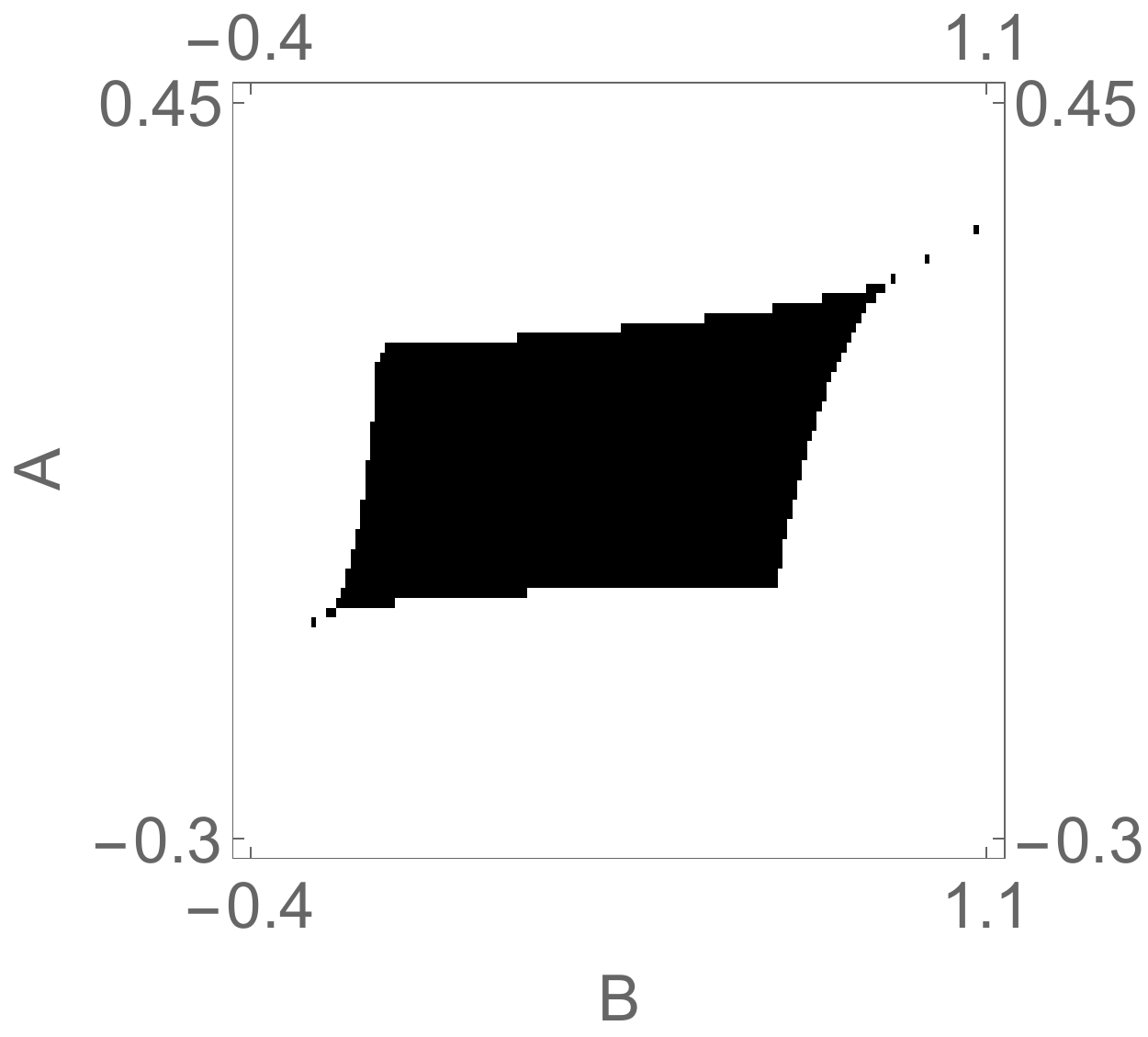}}}
		\subfloat[]{\raisebox{-0.5\height}{\includegraphics[width=.24\linewidth]{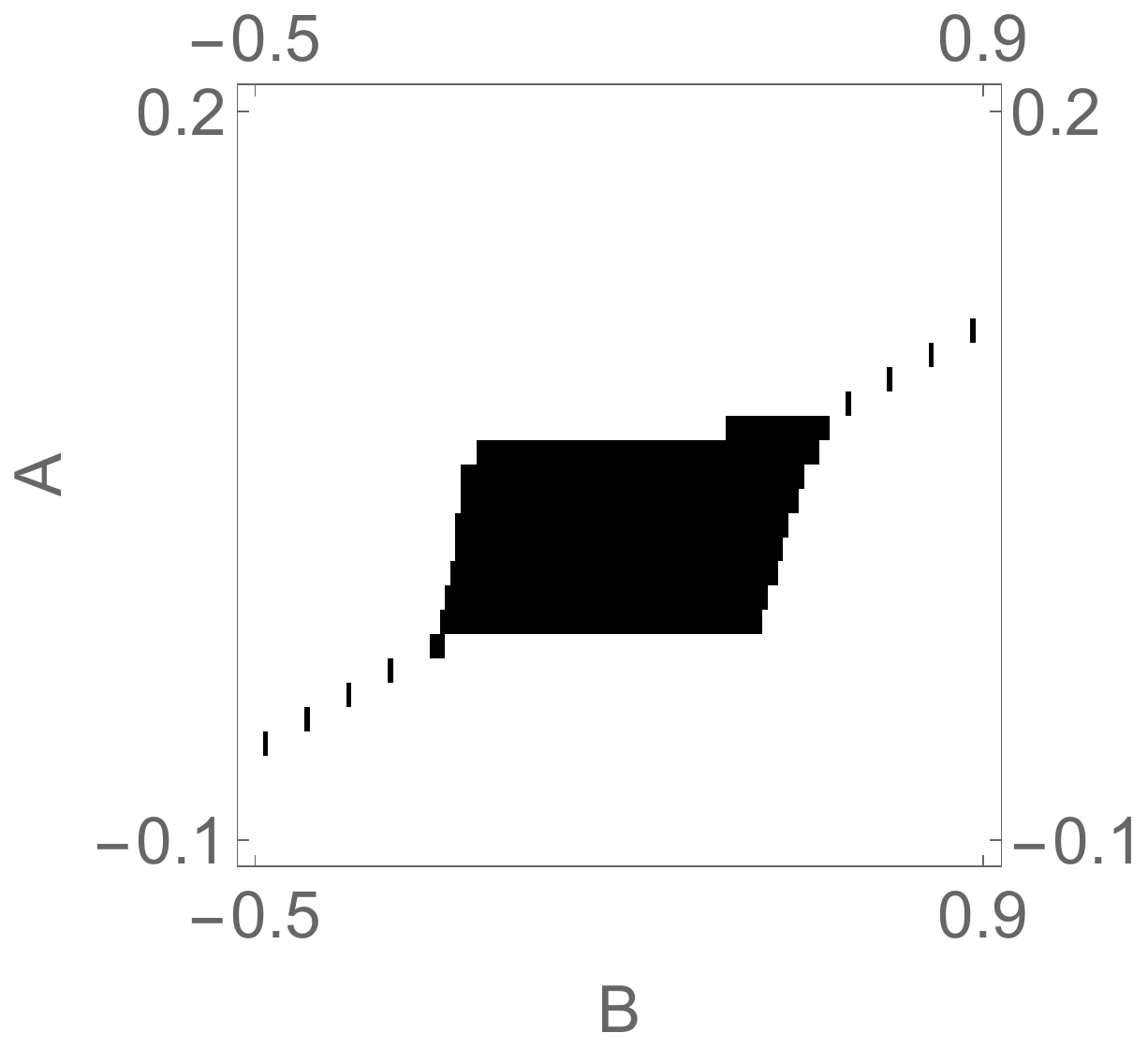}}}
		\subfloat[]{\raisebox{-0.5\height}{\includegraphics[width=.24\linewidth]{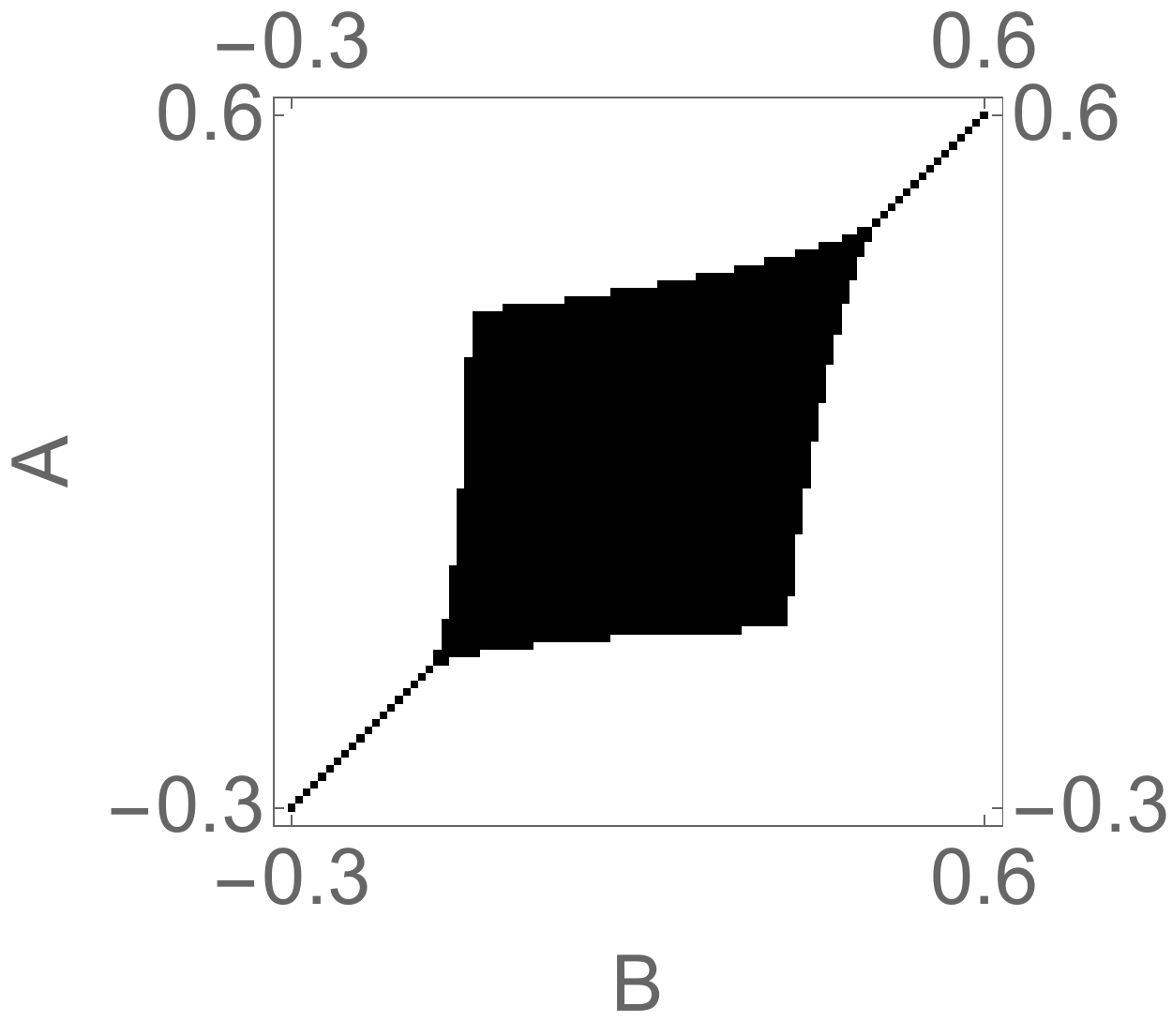}}}\\
		\subfloat[]{\raisebox{-0.5\height}{\includegraphics[width=.24\linewidth]{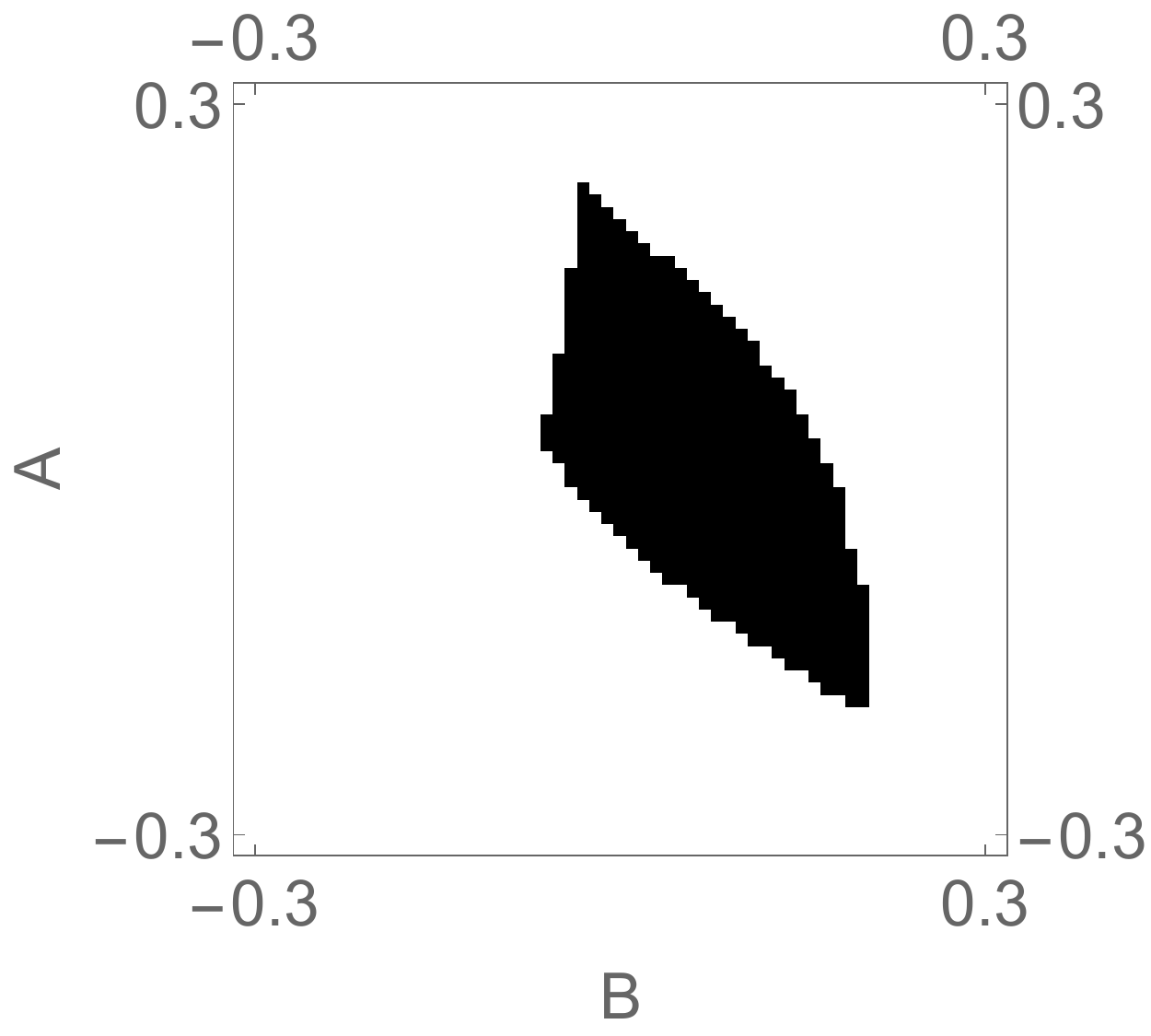}}}
		\subfloat[]{\raisebox{-0.5\height}{\includegraphics[width=.24\linewidth]{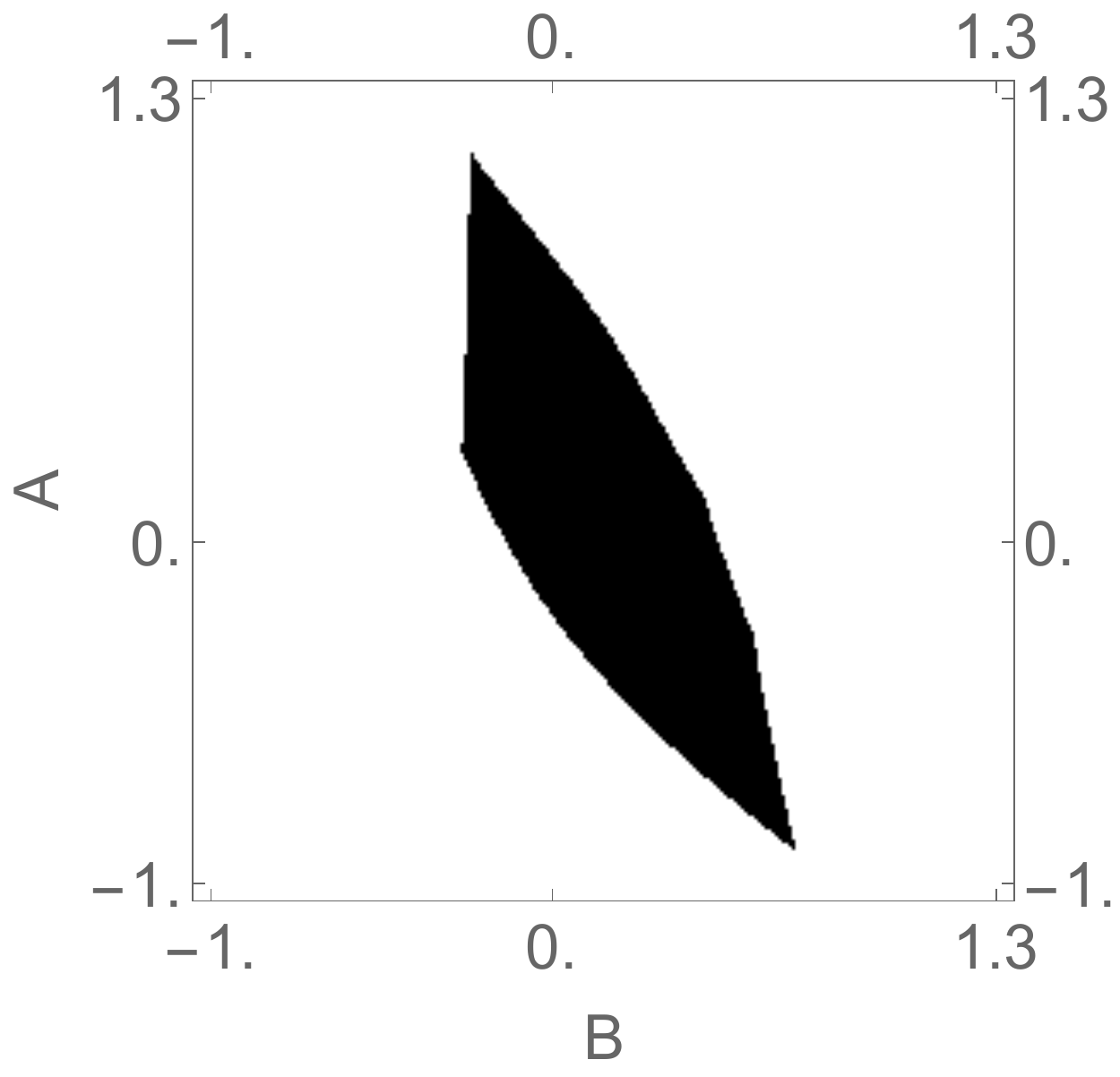}}}
		\subfloat[]{\raisebox{-0.5\height}{\includegraphics[width=.24\linewidth]{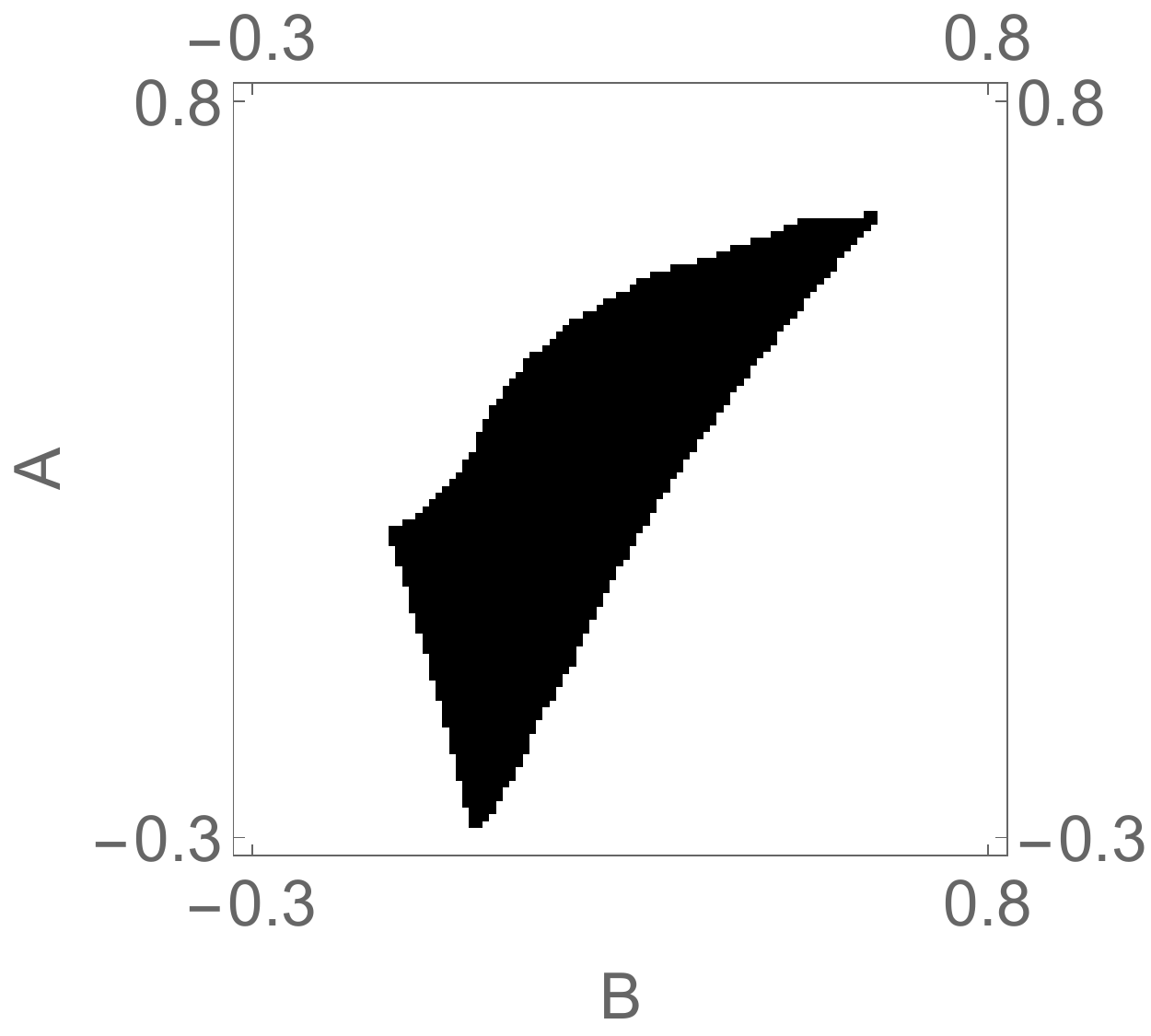}}}
		\subfloat[]{\raisebox{-0.5\height}{\includegraphics[width=.24\linewidth]{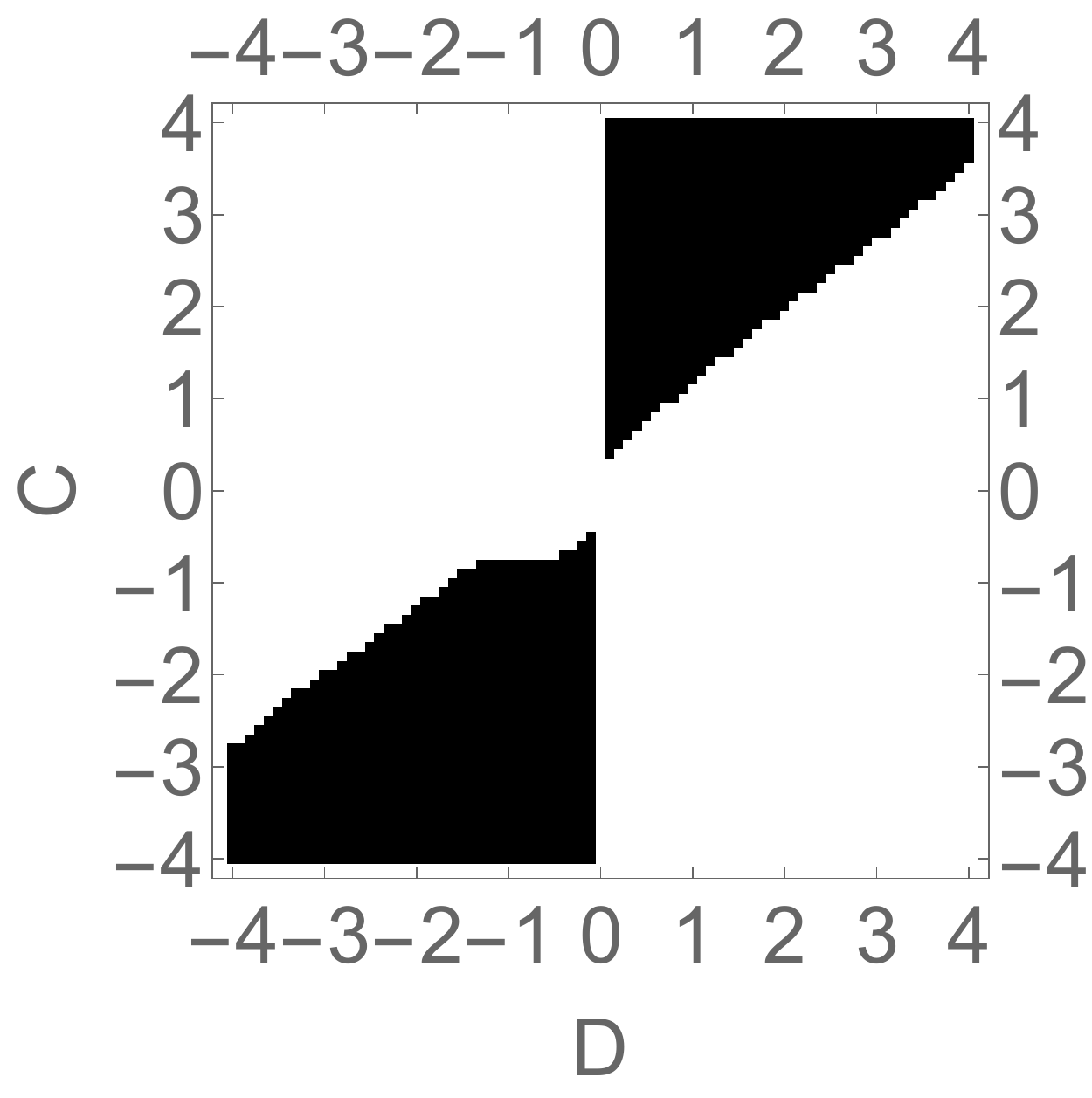}}}
		\centering
		\caption{
			Regions of the parameter space with real OBC spectra (black) for the following 2-component separable models: (a) $H_\text{2-band}^0$; (b) $H_\text{2-band}^1$ with $C=2,D=3$. (c) $H_\text{2-band}^1$ with $C=1/2,D=1$; (d) $H_\text{2-band}^1$ with $C=2,D=2$; (e) $H_\text{2-band}^2$ with $C=2,D=3$;	(f) $H_\text{2-band}^2$ with $C=2,D=1$; (g) $H_\text{2-band}^3$ with $C=2,D=2$; (h) $H_\text{2-band}^3$ with $A=0.1,B=0.1$. The numerical threshold is  $\text{Max}|\text{Im}(E)|<\epsilon=10^{-6}$. Note that the PBC spectra of all of these models are complex. }
		\label{fig:3} 
	\end{figure*}
	
	In this case of constant trace, the dispersion is essentially determined by $F(E)=-G(p)=\text{Det}H_2(p)$, directly generalizing the case of 1-component models upon the conformal transform $E\rightarrow E^2-[\text{Tr}H_2]E$. The only difference is that $G(p)$ now contains products of matrix elements of $H_2(p)$, and as such is usually a higher-order polynomial describing OBC spectra with no analytic solution. While a higher-order $G(p)$ can result in a more branched and hence complex OBC spectrum, with appropriate model design, there can still be larger parameter regions with real OBC spectra, see Fig.~\ref{fig:3}. 
	
	For instructive purposes, we first study the 2-component model with almost trivial NHSE:
	\begin{equation}
		H_\text{2-band}^0(z)=\left(z+Az^{-1}\right)\sigma_x+B\sigma_y,
	\end{equation}
	$\sigma_x,\sigma_y$ the Pauli matrices. Upon performing the $\kappa_0$ translation $p\rightarrow p+i\kappa_0=p+i\log\sqrt{A}$, we find the dispersion $E^2=-G(p)=B^2+2A\cos 2p$. As such, the spectrum is real whenever $B^2>-2A$, as numerically verified in Fig.~\ref{fig:3}a.
	
	Next, inspired by the 1-component cases, we present the following models with nontrivial real spectral parameter regions:
	\begin{equation}
		H_\text{2-band}^1(z)=\left(Az^2+\frac1{z}+C\right)\sigma_++\left(z+\frac{B}{z^2}+D\right)\sigma_-,
	\end{equation}
	\begin{equation}
		H_\text{2-band}^2(z)=\left(Az^2+\frac1{z}+C\right)\sigma_++\left(z+Bz^3+D\right)\sigma_-
	\end{equation}
	\begin{equation}
		H_\text{2-band}^3(z)=\left(\frac{A}{z^2}+\frac1{z}+C\right)\sigma_++\left(z+\frac{B}{z^3}+D\right)\sigma_-
	\end{equation}
	where $\sigma_\pm=(\sigma_x\pm i\sigma_y)/2$ and $z=e^{ip}$. In these models, the hoppings are different in either direction, and it is surprising from Fig.~\ref{fig:3}b-h that real OBC spectra not just exists, but in fact over large parameter space regions. For instance, in $H^2_\text{2-band}$ with $C=2,D=1$ (Fig.~\ref{fig:3}h), the OBC spectrum can be real even for $A>1$, which corresponds to a very unbalanced set of physical couplings.

	\subsection{Inseparable energy dispersions $P(E,p)$}
	
	We next consider more sophisticated dispersions which contain products of $E$ and $z=e^{ik}$. In general, they are not analytically tractable, although their spectral graph structure can often still be heuristically predicted~\cite{tai2022zoology}. In this work, we shall limit ourselves to 2-component models. From Eq.~\ref{P2Ep}, those with inseparable dispersions correspond to those with $p$-dependent traces, i.e. those with same-sublattice hoppings across unit cells.
	
	\subsubsection{Analytically tractable examples}
	
\begin{figure*}
		\subfloat[]{\raisebox{-0.5\height}{\includegraphics[width=.26\linewidth]{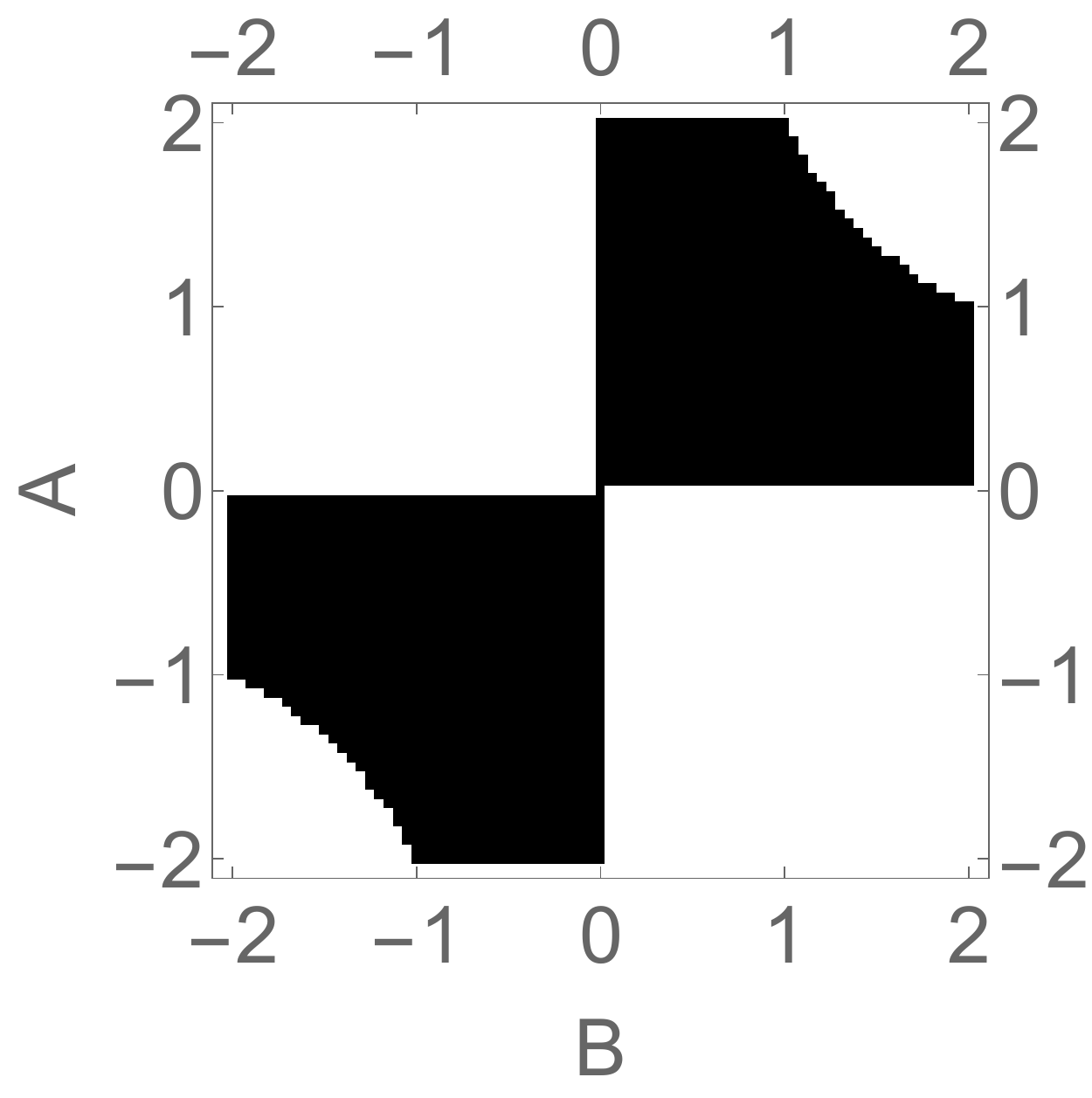}}}
		\subfloat[]{\raisebox{-0.5\height}{\includegraphics[width=.26\linewidth]{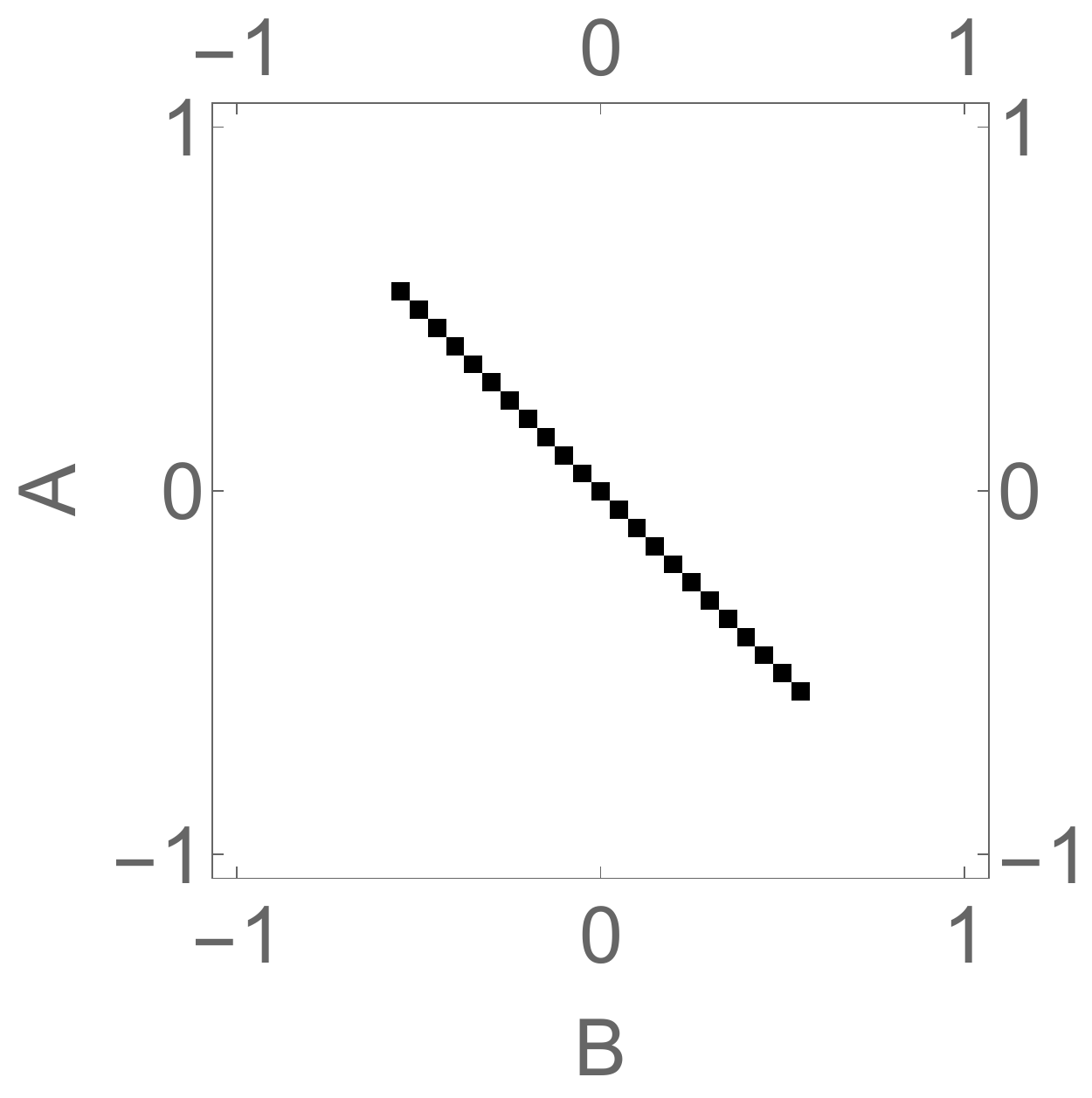}}}
		\subfloat[]{\raisebox{-0.5\height}{\includegraphics[width=.26\linewidth]{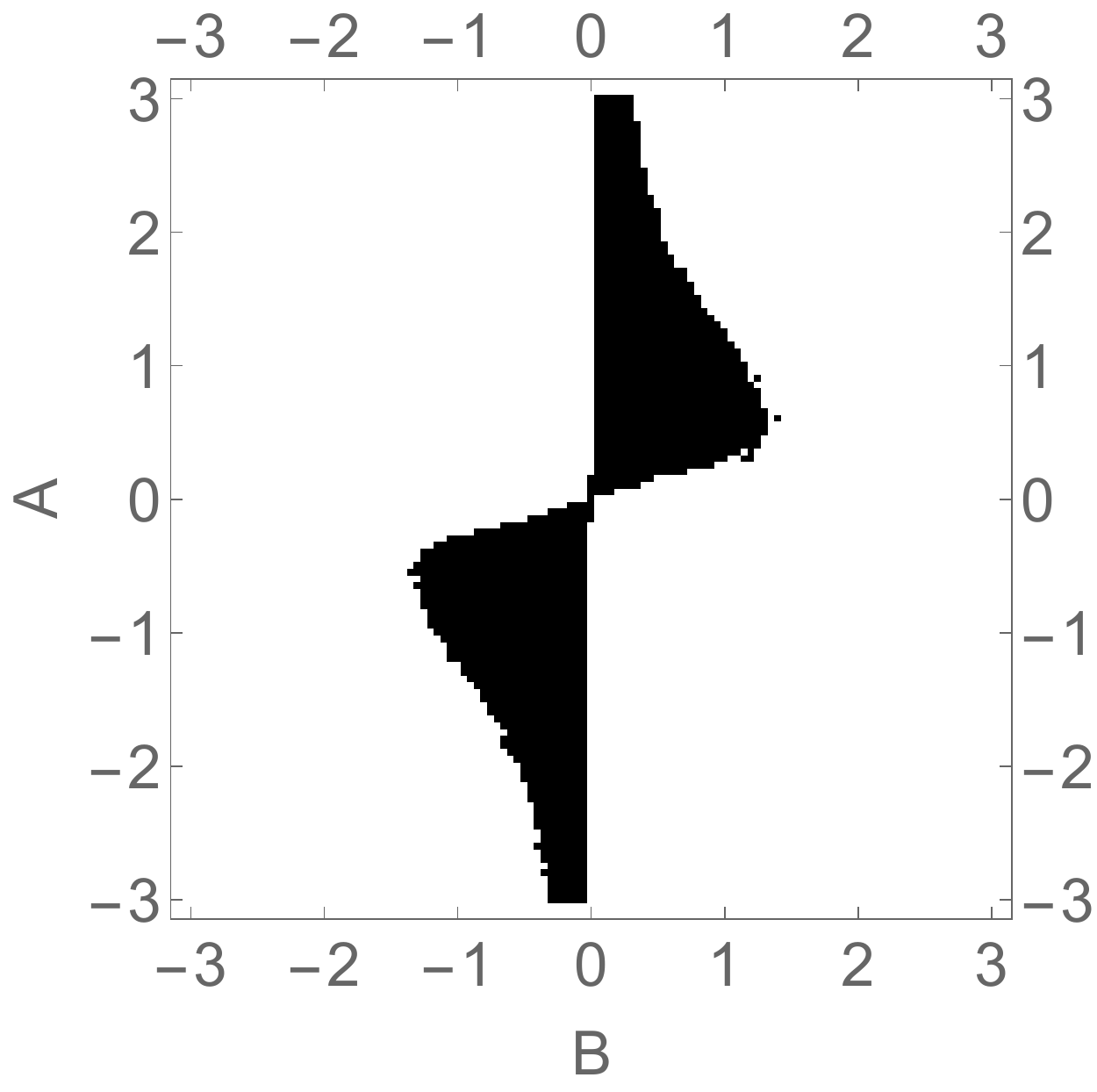}}}\\
		\subfloat[]{\raisebox{-0.5\height}{\includegraphics[width=.26\linewidth]{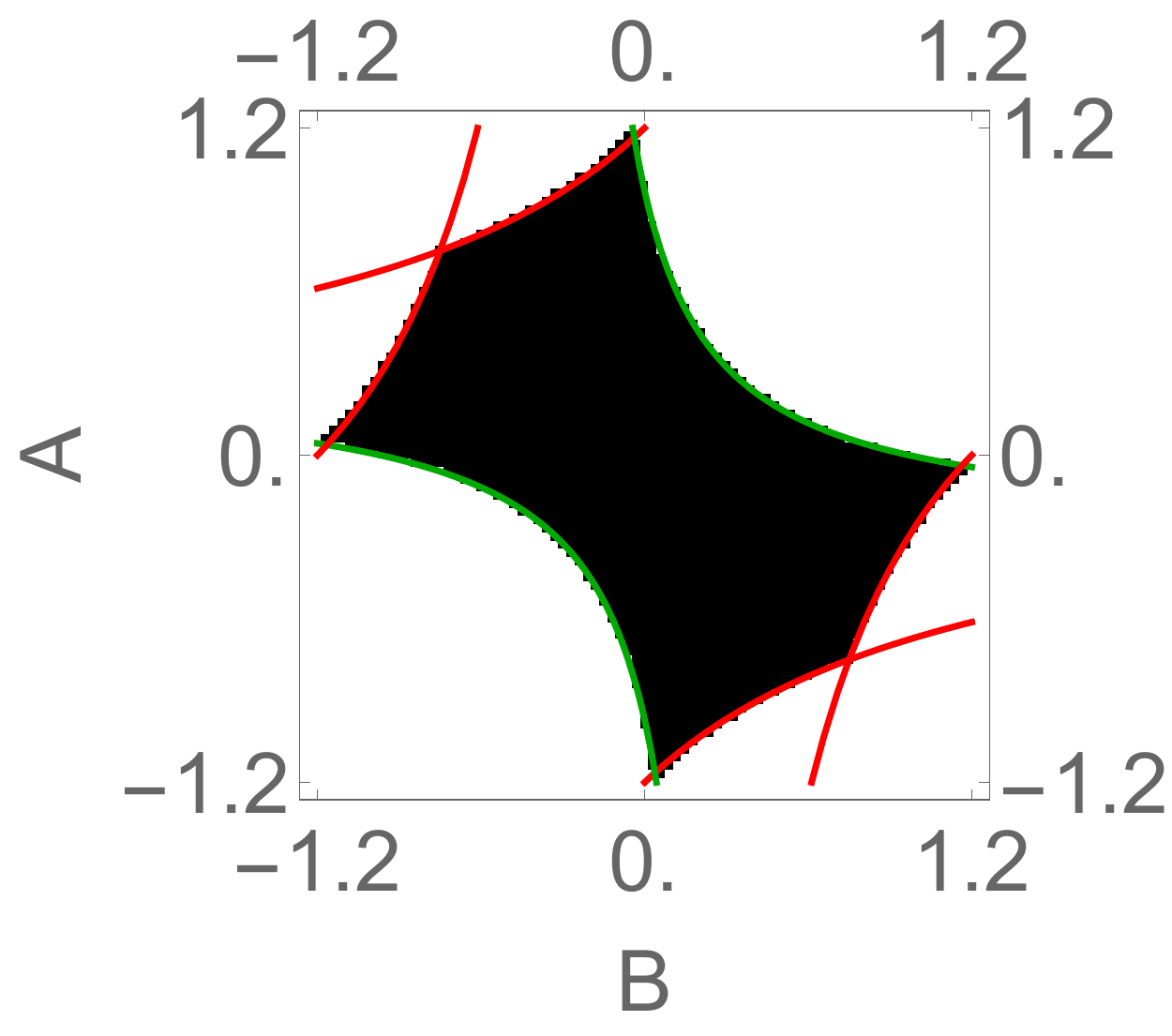}}}
		\subfloat[]{\raisebox{-0.5\height}{\includegraphics[width=.26\linewidth]{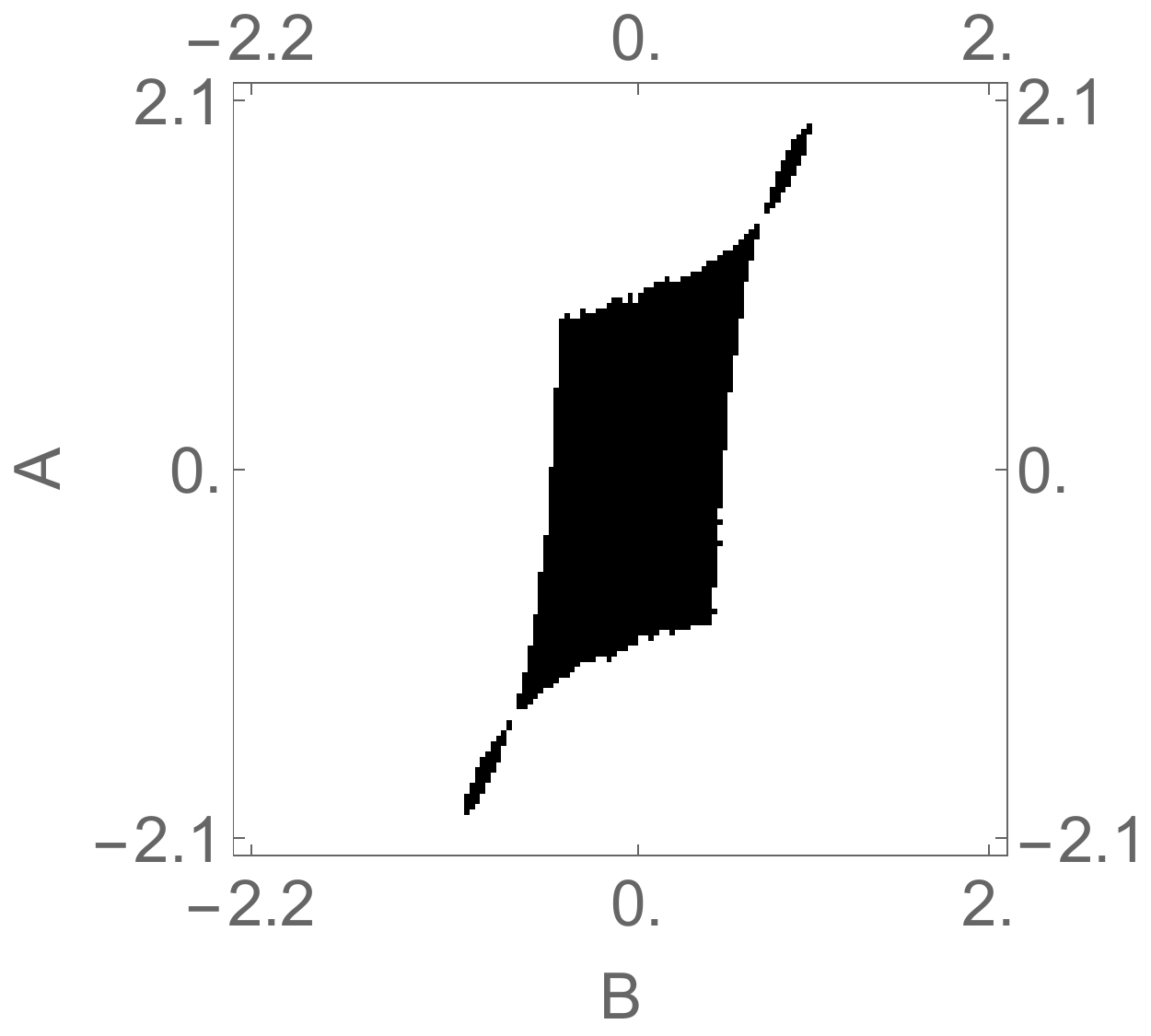}}}
		\subfloat[]{\raisebox{-0.5\height}{\includegraphics[width=.26\linewidth]{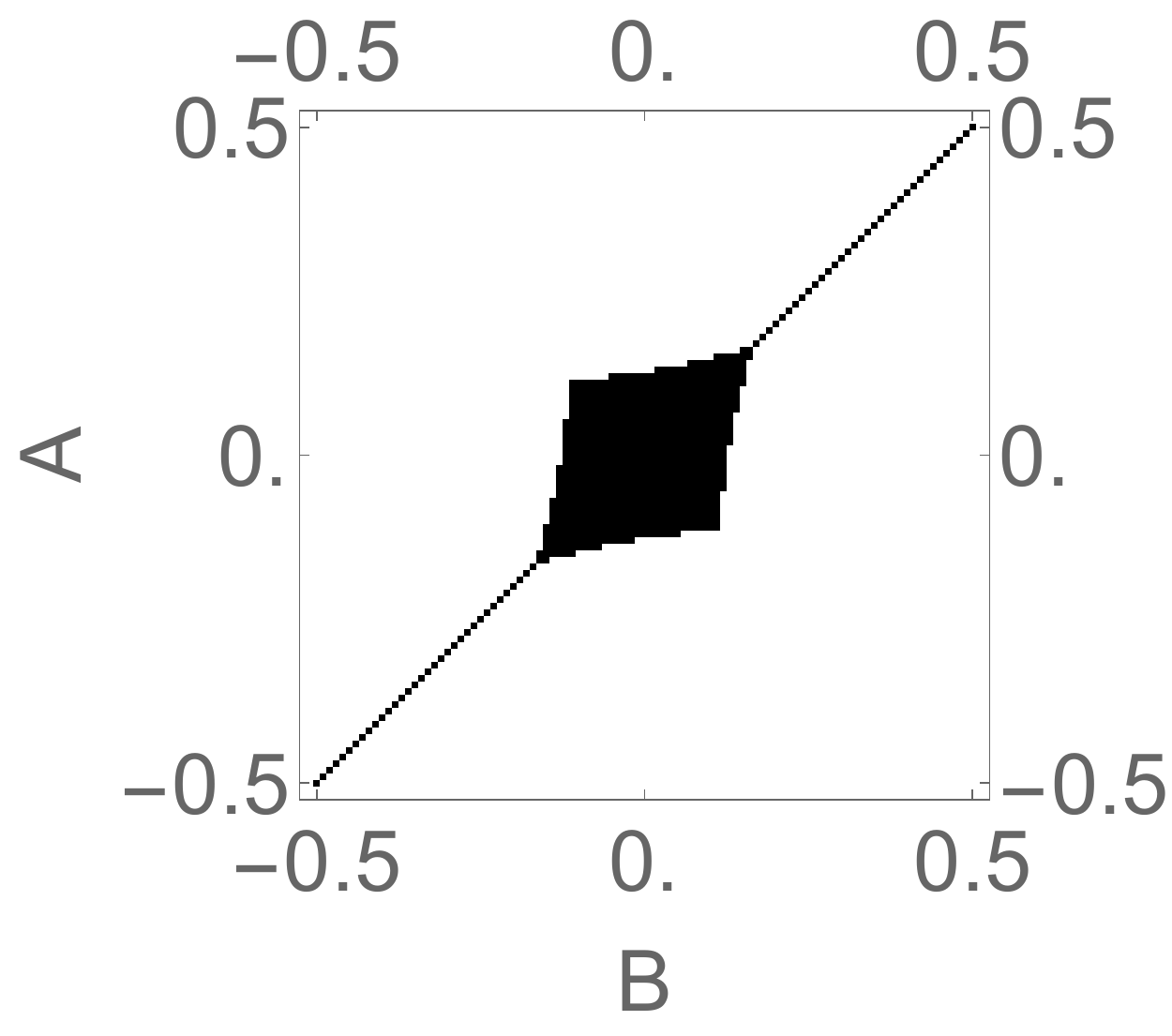}}}
		\caption{Regions of the parameter space with real OBC spectra (black) for the following 2-component inseparable models: (a) $H_\text{in}^1(z)$; (b) $H_\text{in}^2(z)$; (c) $H_\text{in}^3(z)$; (d) $H_\text{in}^4(z)$ with the regime boundary approximately depicted by curves $(0.28+A)(0.28+B)=0.35$ and its reflection across line $A=-B$ (green), $1.2=A(-0.8B+1)$ and its reflection across line $A=-B$ plus further reflection across line $A=B$ (red); (e) $H_\text{in}^5(z)$; (f) $H_\text{in}^6(z)$. The system length is $L=30$ for all and the numerical threshold is  $\text{Max}|\text{Im}(E)|<\epsilon$ where $\epsilon=10^{-5}$ for (c), and $\epsilon=10^{-6}$ for the other cases. }
		\label{fig:4} 
	\end{figure*}

	First, we introduce an inseparable case whose condition for real spectrum can still be analytically derived. Consider
	\begin{equation}
		H_\text{in}^1(z)=\left(\begin{matrix}
			Az^3 & \sqrt{C} \\ \sqrt{C} & B/z^3
		\end{matrix}\right)
		\label{Hin1}
	\end{equation}
	with dispersion given by the characteristic polynomial
	\begin{equation}
		P_\text{in}^1(E,z)=E^2-\left(Az^3+\frac{B}{z^3}\right)E+AB-C
	\end{equation}
	where $z=e^{ip}$. To make analytic headway, we perform the hopping rescaling $p\rightarrow p+i\log \sqrt[6]{A/B}$ so as to symmetrize the $Az^3+Bz^{-3}$ term to $\sqrt{AB}(z^3+z^{-3})$. Doing so does not change the OBC spectrum; yet, due to the symmetric occurrence of $z$ and $z^{-1}$, we also know that $\kappa=0$ i.e. real $p$ constitutes a potential solution for the OBC spectrum. Whether real $p$ indeed generates \emph{the correct solution} depends on whether a competing degenerate $\kappa$ solution exists. For this model, it is not hard to show that, with $p\rightarrow i\log \sqrt[6]{B/A}$, 
	\begin{eqnarray}
		P_\text{in}^1(E,z)&=&E^2-2\sqrt{AB}E\cos 3p+AB-C\notag\\
		&=& \left(E-\sqrt{AB}\cos 3p\right )^2+AB\sin^23p-C=0\qquad
	\end{eqnarray}
	admits real $E$ solutions generated by real $p$ as long as $0\leq AB\leq C$. Outside of this regime, complex pairs of $E$ solutions appear (Fig.~\ref{fig:4}a), leading to non-real spectra. Fig.~\ref{fig:5} shows the boundary behaviour of spectra from real to non-real as the parameter varies. This criterion $0\leq AB\leq C$ entails that the real spectrum hinges on the presence of nonzero coupling, and is numerically verified in Fig.~\ref{fig:4}a. 
	
	Next, we introduce another Hamiltonian which admits real spectrum for parameters that make it separable:
	\begin{equation}
		H_\text{in}^2(z)=\left(\begin{matrix}
			Az & 1+z^2 \\ 2+1/z^2 & Bz
		\end{matrix}\right)
		\label{Hin2}
	\end{equation}
	with 
	\begin{equation}
		P_\text{in}^2(E,z)=E^2-(A+B)Ez+(AB-2)z^2-3-z^{-2}.
	\end{equation}
	Although its GBZ and thus OBC spectrum is not analytically solvable for general parameters, for the special case when $A=-B$, the trace term disappears, and we simply obtain $E=\pm \sqrt{(2+A^2)z^2+z^{-2}+3}$. Again, this is of Hatano-Nelson form, and setting $p\rightarrow p+i\log\sqrt[4]{2+A^2}$, we obtain $\bar E =\pm\sqrt{3+2\sqrt{2+A^2}\cos 2p}$. This is real for $A=-B$ with $|A|<1/2$, as reflected in its numerical parameter space diagram (Fig.~\ref{fig:4}b).
	
	\subsubsection{More general examples}
	
	Next, we introduce a few more models whose propensity for real OBC spectra cannot be predicted through any simple way:
	\begin{equation}
		H_\text{in}^3(z)=\left(\begin{matrix}
			2+Az & 2 \\ 1 & 1+B/z
		\end{matrix}\right)
		\label{Hin3}
	\end{equation}
	\begin{equation}
		H_\text{in}^4(z)=\left(\begin{matrix}
			1+Az & 1+z \\ 2+1/z & 1+Bz
		\end{matrix}\right)
		\label{Hin4}
	\end{equation}
	\begin{equation}
		H_\text{in}^5(z)=\left(\begin{matrix}
			Az & 1+z \\ 2+1/z & B/z
		\end{matrix}\right)
		\label{Hin5}
	\end{equation}
	\begin{equation}
		H_\text{in}^6(z)=\left(\begin{matrix}
			z+Az^2 & 2 \\ 1 & 1/z+B/z^2
		\end{matrix}\right)
		\label{Hin6}
	\end{equation}
	Like $H_\text{in}^1$ of Eq.~\ref{Hin1}, the model $H_\text{in}^3$ contain constant off-diagonal couplings. However, despite its simple algebraic form, it is not analytically tractable, and in fact behaviors completely differently from $H_\text{in}^1$, and in fact the other models too. 
	
	The real spectrum parameter region of $H_\text{in}^6$ is an interesting intersection of a diagonal $A=B$ line segment, and a smaller extended region. These two parameter subregions have different origins: For the former, the OBC spectrum is real because the PBC spectrum is also real; larger $A=B$ closes the band gap and releases complex eigenenergies. For the latter, the PBC spectrum is always complex (example in Fig.~\ref{fig:6}), but an open boundary creates sufficient interference to stop indefinite amplification, leading to a completely real spectrum.

	\section{Discussion}
	
	In this work, we have seen that real non-Hermitian OBC spectra are generically more robust than real PBC spectra. This can be explained in terms of the inverse skin depth $\kappa(E)$ solutions curves - while the PBC spectrum becomes complex once $\kappa(E)\neq 0$ at $\text{Im}E \neq 0$, a complex OBC spectrum requires $\kappa(E)$ curves to \emph{intersect} at $\text{Im}E \neq 0$, which is a more demanding condition. While there exists a very general electrostatics approach~\cite{yang2022designing} that returns possible parent Hamiltonians for any desired real OBC spectrum and skin localization, this work showcases particularly simple ansatz models that have the benefit of being as local as possible.
	
	The discovery of these Hamiltonians with real spectra complements existing efforts towards the design of stable non-Hermitian system, particularly the PT-symmetry route. With a greater set of ansatze models that are not symmetry constrained, a larger variety of interesting non-Hermitian physics i.e. non-Hermitian skin clusters and pseudogaps~\cite{li2022non,shen2021nonhermitian,tahir2011PG} can be realized not just with enhanced stability, but also with generically non-reciprocal platforms such as circuits with MOSFETs or operational amplifiers~\cite{budich2020sensor,yang2022observation,hofmann2019chiral,helbig2020generalized,zou2021observation,ningyuan2015time,stegmaier2021topological,blais2021circuit}. Yet, the NHSE is not guaranteed to yield real spectra, and in Appendix A, we have also listed down models which failed, to aid further search efforts.

	\bibliographystyle{ieeetr}
	\bibliography{referencesQY}
	
	\clearpage
	\appendix
	
	\section{Inseparable energy dispersions not giving real spectra}

	While the main text had discussed various models with real spectra despite having no favorable symmetries, it is also important to record models where this is does not occur. As discussed, NHSE-induced real spectra depends on the algebraic properties of the $\kappa(E)$ curves, and it is instructive to list the models where they do not behave favorably as such.
	
	For instance, the following deformations of $H_\text{in}^1(z)$ have no real spectra:
	\begin{itemize}
		\item $	H_\text{in,m1}^1(z)=\left(\begin{matrix}
			Az^{n} & 0 \\ c_{1} & Bz^{-n}
		\end{matrix}\right)$ and
		\item $ H_\text{in,m2}^1(z)=\left(\begin{matrix}
			Az^{n} & -c_{1} \\ c_{1} & Bz^{-n}
		\end{matrix}\right),$
		for sufficiently small $c_1$, which is expected as they do not fulfil the condition of $0\leq AB\leq C$, where $C=0$ for $H_\text{in,m1}^1$ and $C=c_{1}^2$ for $H_\text{in,m2}^1$.
		
		\item	$ H_\text{in,m3}^1(z)=\left(\begin{matrix}
			Az^{n} & 0 \\ c_{1} & Bz^{n}
		\end{matrix}\right).$
	\end{itemize}
	Modifications of $H_\text{in}^2(z)$ and  $H_\text{in}^4(z)$ with diagonal terms having all positive powers of $z$ only:
	\begin{itemize}
		\item $H_\text{in,m1}^4(z)=\left(\begin{matrix}
			1+Az & 1+z \\ 2+z & 1+Bz
		\end{matrix}\right).$
	\end{itemize}
	
	Modifications of $H_\text{in}^3(z)$ with diagonal terms having same sign of power of $z$:
	\begin{itemize}
		\item By having same sign of power of $z$ at the non-diagonal terms\\
		
		$H_\text{in,m1}^3(z)=\left(\begin{matrix}
			2+Az & c_{2} \\ 1 & 1+Bz
		\end{matrix}\right),$
		with $c_{2}=1, -1, 0$.
	\end{itemize}
	Modifications of $H_\text{in}^5(z)$ with diagonal terms having positive and negative powers of $z$:
	\begin{itemize}
		\item 	$H_\text{in,m1}^5(z)=\left(\begin{matrix}
			Az & 1+z^2 \\ 2+z^{-2} & Bz^{-1}
		\end{matrix}\right),$
		\item By having same sign of power of $z$ at the non-diagonal terms\\
		
		$H_\text{in,m2}^5(z)=\left(\begin{matrix}
			Az & 1+z \\ 2+z & Bz^{-1}
		\end{matrix}\right),$
	\end{itemize}
	Modifications of $H_\text{in}^6(z)$ with diagonal terms having same sign of power of $z$:
	\begin{itemize}
		\item 	
		$H_\text{in,m1}^6(z)=\left(\begin{matrix}
			1+z+Az^2 & 1 \\ 2 & 1+z+Bz^{-2}
		\end{matrix}\right).$\\
		$H_\text{in,m2}^6(z)=\left(\begin{matrix}
			1+z+Az^2 & 1 \\ 2 & 1+z^{-1}+Bz^{2}
		\end{matrix}\right).$
		
	\end{itemize}
	Models with non-symmetric hopping systems:
	\begin{itemize}
		\item 
		$H_\text{c1}(z)=\left(\begin{matrix}
			Az^2 & z \\ 2 & Bz^{-2}
		\end{matrix}\right).$
		\item 	$H_\text{c2}(z)=\left(\begin{matrix}
			Az^2 & z \\ 2 & Bz^{2}
		\end{matrix}\right).$
		\item 
		$H_\text{c3}(z)=\left(\begin{matrix}
			Az^2 & z+2 \\ 2 & Bz^{-2}
		\end{matrix}\right).$
		\item 
		$H_\text{c4}(z)=\left(\begin{matrix}
			Az^2 & z+2 \\ 2 & z^{-1}+Bz^{-2}
		\end{matrix}\right).$
		\item 
		$H_\text{c5}(z)=\left(\begin{matrix}
			1+z+Az^2 & 2 \\ 2 & Bz^{-2}
		\end{matrix}\right).$
	\end{itemize}

	\section{Spectra of illustrative models}
	
	Here, we display the spectra of selected models discussed, such as to contrast the OBC vs. PBC spectra and their propensities for being real.

	\begin{figure}[H]
		\includegraphics[width=.9\linewidth]{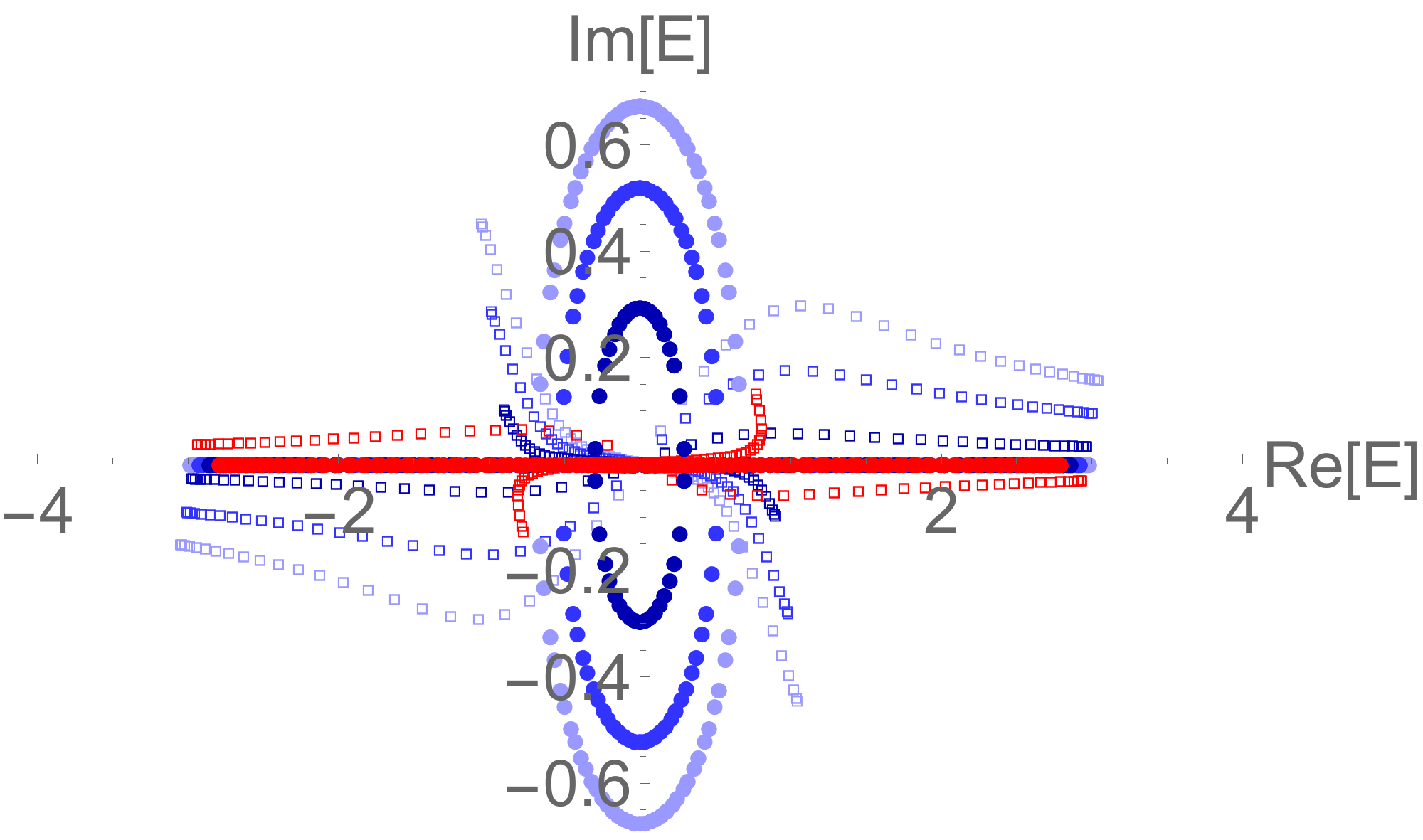}
		\caption{
			Spectra of model $H_\text{in}^1$ (Eq.~\ref{Hin1}) for parameters $B=0.9$ and $A=1$ (red real spectrum), $A=1.1, 1.2, 1.3$ (decaying blue complex spectra). Dots/square represents OBC/PBC eigenenergies.  The OBC spectrum becomes real as PBC spectrum evolves across the real line.\\\hspace{\textwidth}
		}
		\label{fig:5}
	\end{figure}

	\begin{figure}[H]
		\subfloat[]{	\includegraphics[width=.49\linewidth]{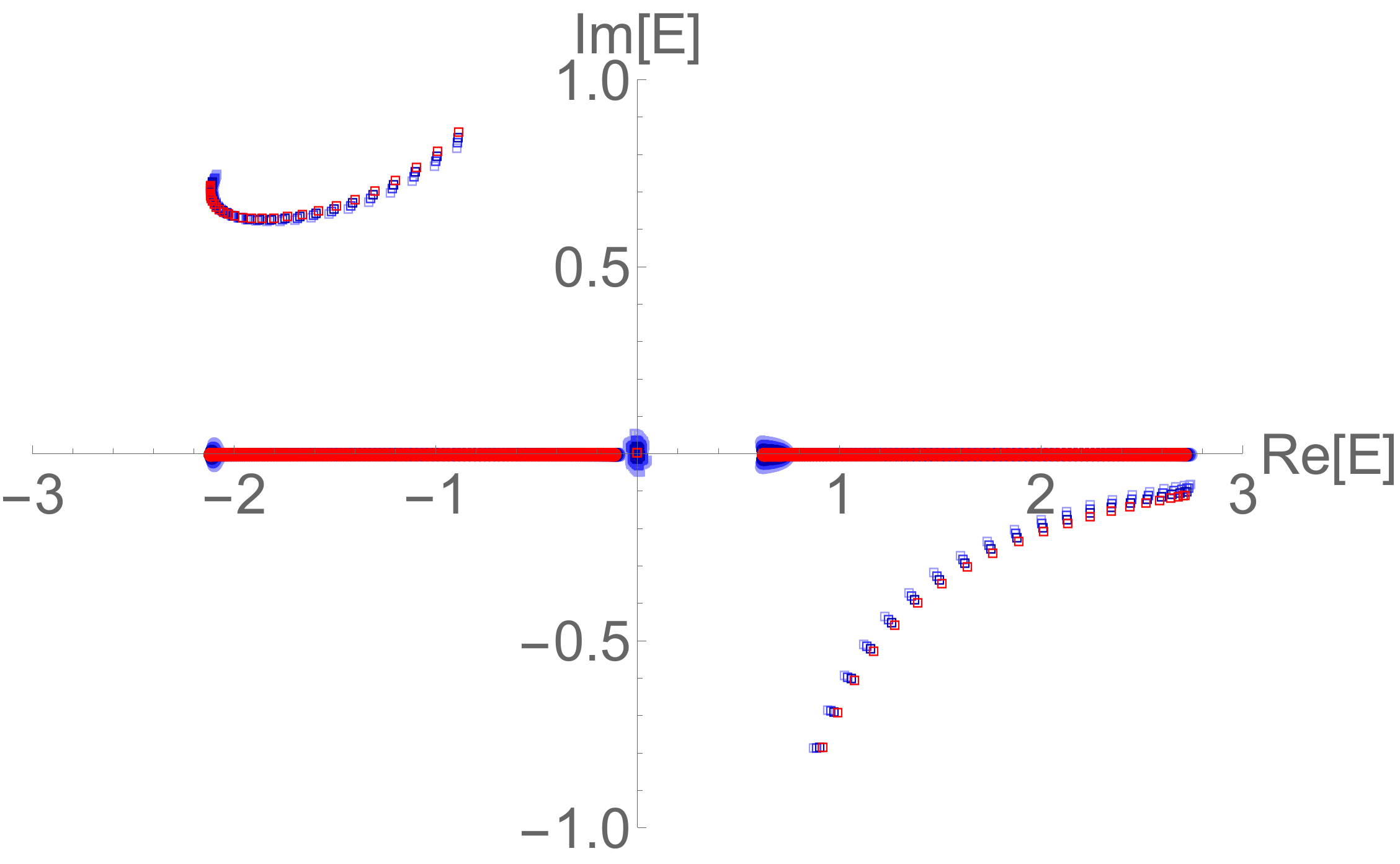}}
		\subfloat[]{	\includegraphics[width=.49\linewidth]{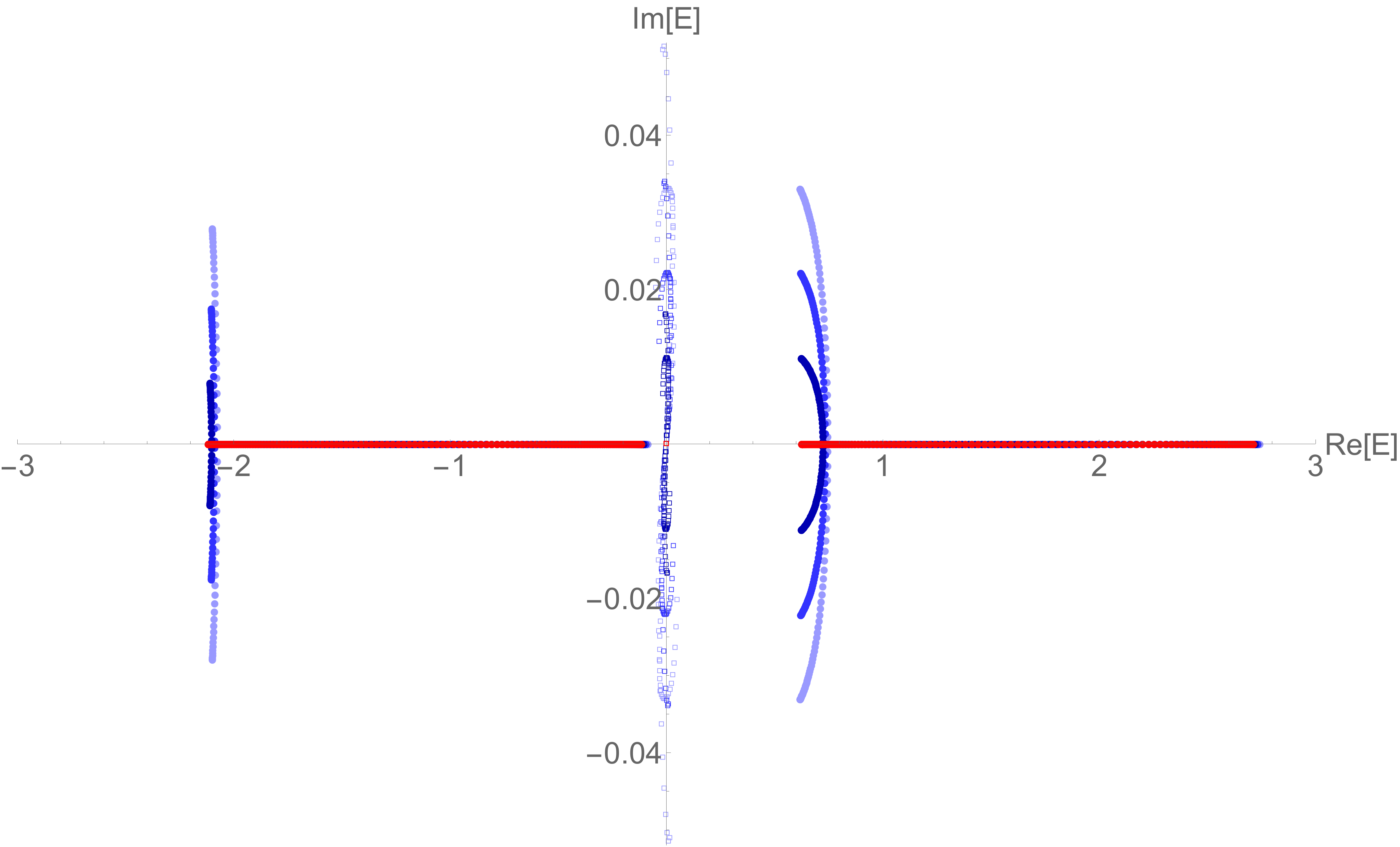}}
		\caption{
			(a) Spectra and (b) Zoomed in spectra of model $H_\text{in}^6$ (Eq.~\ref{Hin6}) for parameters $B=0.3$, $A=0.3$ (red real spectrum) and $A=0.32, 0.34, 0.36$ (decaying blue complex spectra). Dots/square represents OBC/PBC eigenenergies.}
		\label{fig:6}
	\end{figure}

	\begin{figure}[H]
		\includegraphics[width=.9\linewidth]{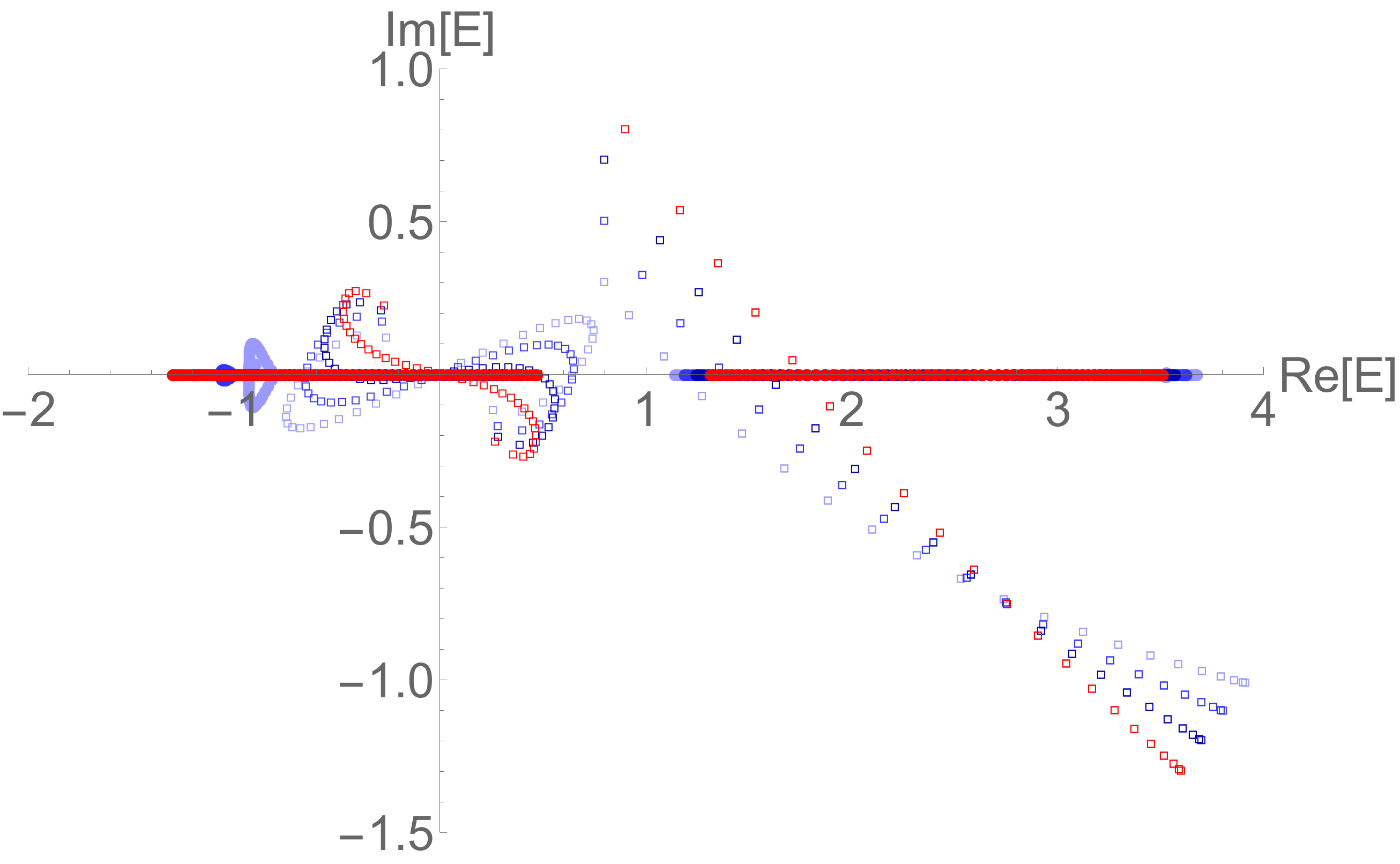}
		\caption{
			Spectra of model $H_\text{in}^4$ (Eq.~\ref{Hin4}) for parameters $B=0.2$, $A=0.1$ (red) and $A=0.3, 0.5, 0.7$ (decaying blue complex spectra). Dots/square represents OBC/PBC eigenenergies.The OBC spectrum becomes real as the near-zero PBC spectrum evolves across the real line.}
		\label{fig:7}
	\end{figure}
	
	\begin{figure}[H]
		\includegraphics[width=.9\linewidth]{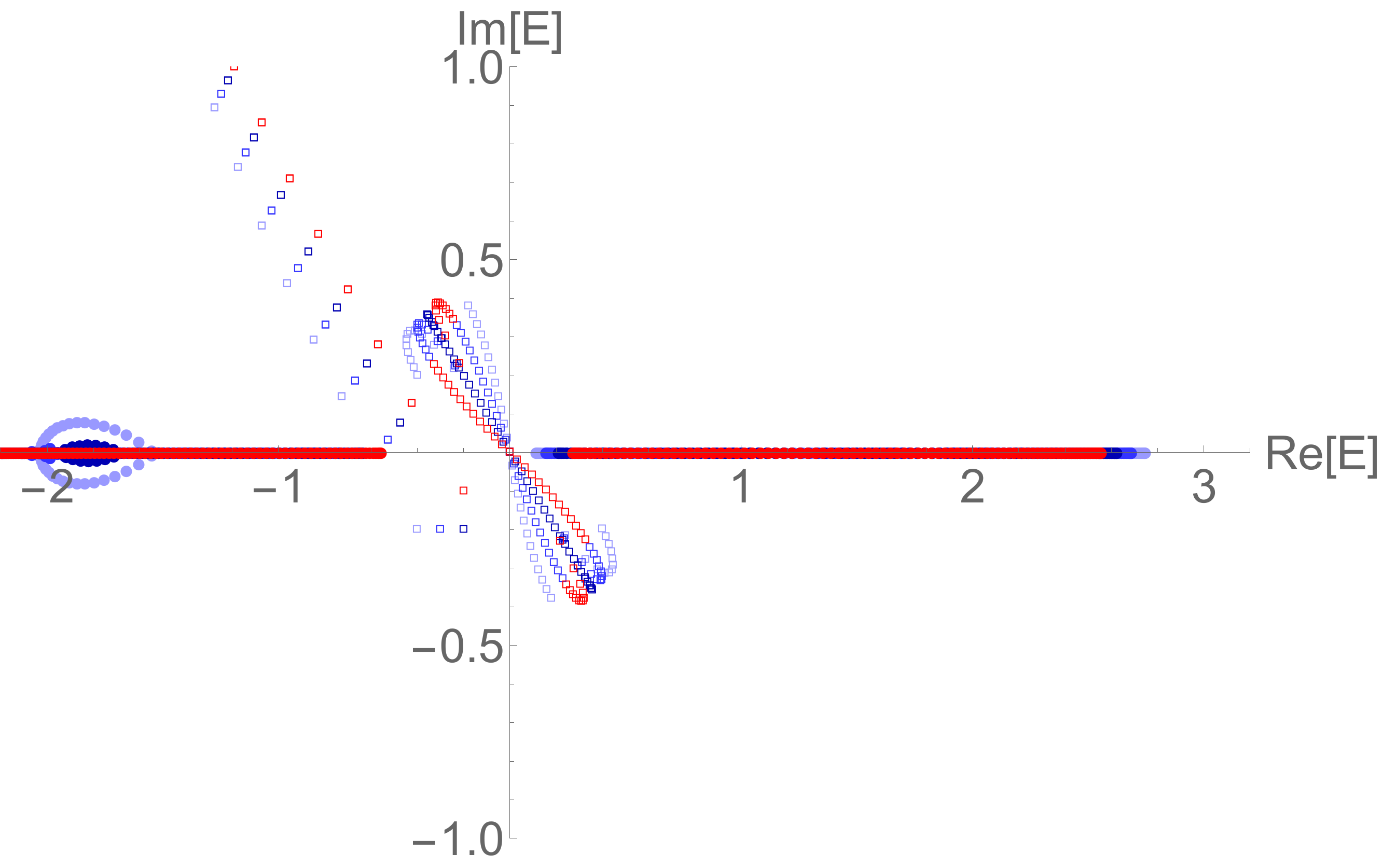}
		\caption{
			Spectra of model $H_\text{in}^5$ (Eq.~\ref{Hin5}) for parameters $B=0.2$, $A=0.1$ (red) and $A=0.2, 0.3, 0.4$ (decaying blue complex spectra). Dots/square represents OBC/PBC eigenenergies.}
		\label{fig:8}
	\end{figure}
	
	\begin{figure}
		\includegraphics[width=.9\linewidth]{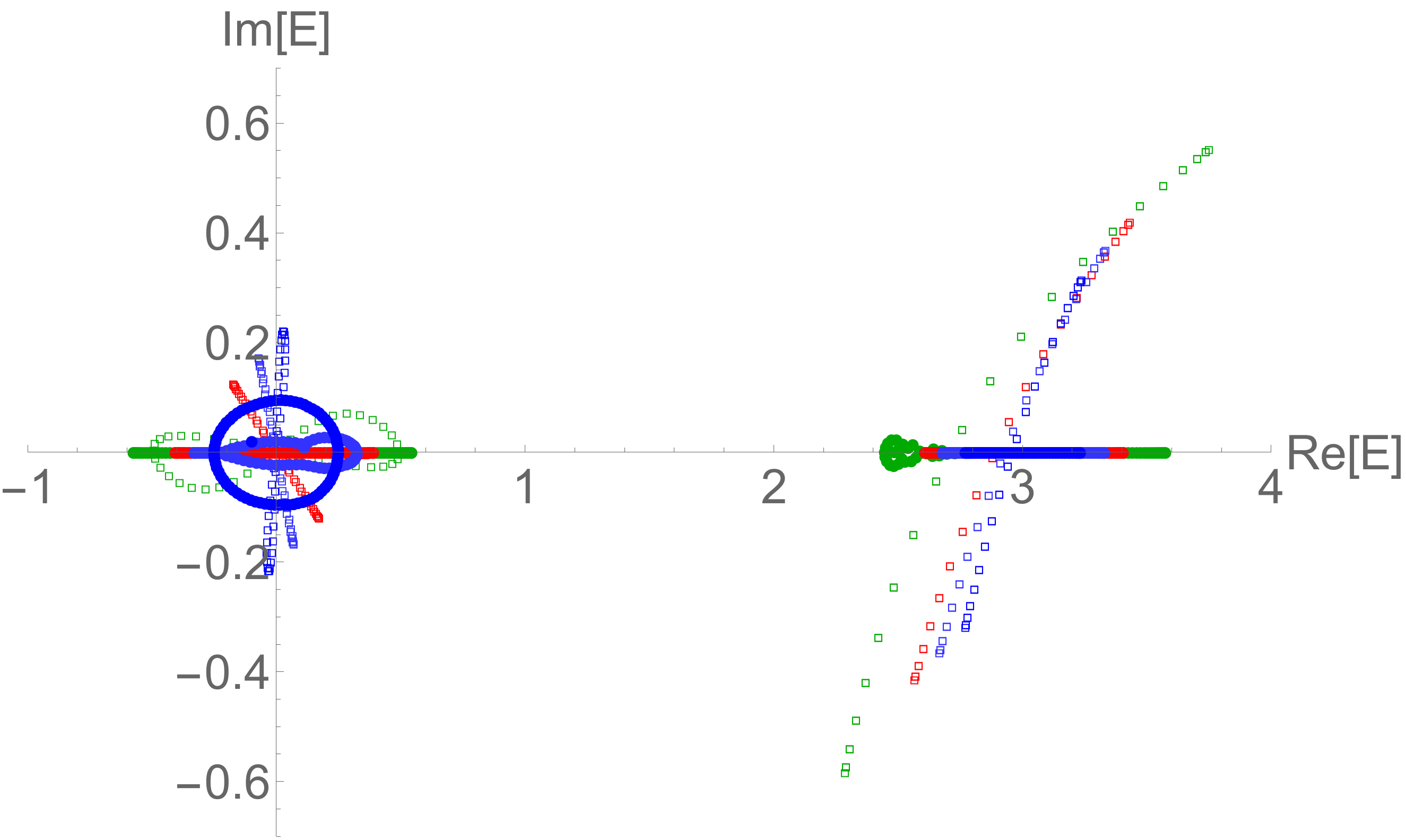}
		\caption{
			Spectra of model $H_\text{in}^3$ (Eq.~\ref{Hin3}) for parameters $B=0.4$ with $A=0.15, 0.3$ (decaying blue complex spectra), $A=0.45$ (Red, real OBC spectrum) and $A=0.9$ (Green). Dots/square represents OBC/PBC eigenenergies. As value of A increases, the complex "bubble" at left side disappears and rises at the right. In between of these two states, a real spectrum shortly appears.}
		\label{fig:9}
	\end{figure}
	
	\begin{widetext}
		
		\begin{figure}
			\subfloat[]{\includegraphics[width=.32\linewidth]{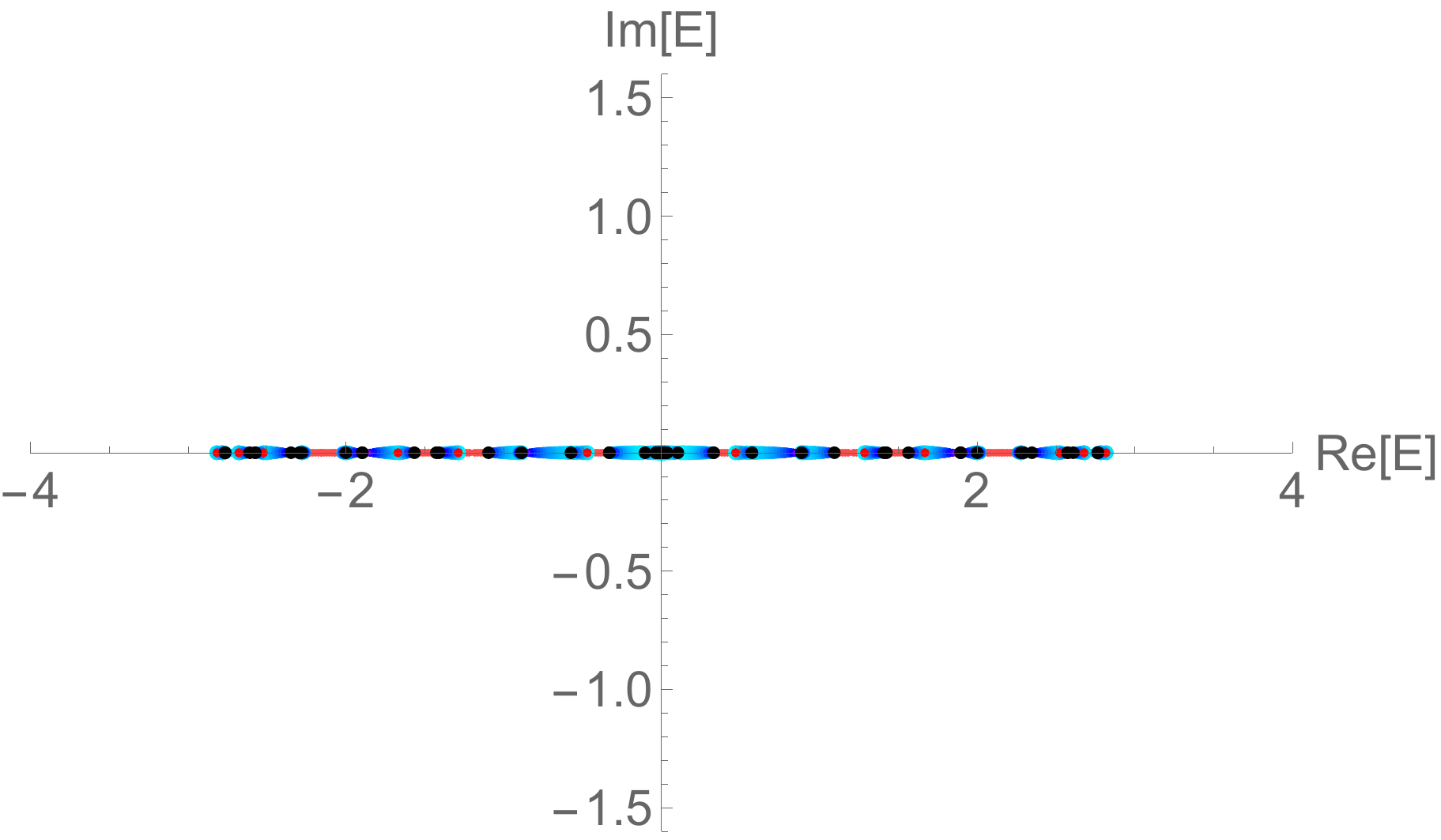}}
			\subfloat[]{\includegraphics[width=.32\linewidth]{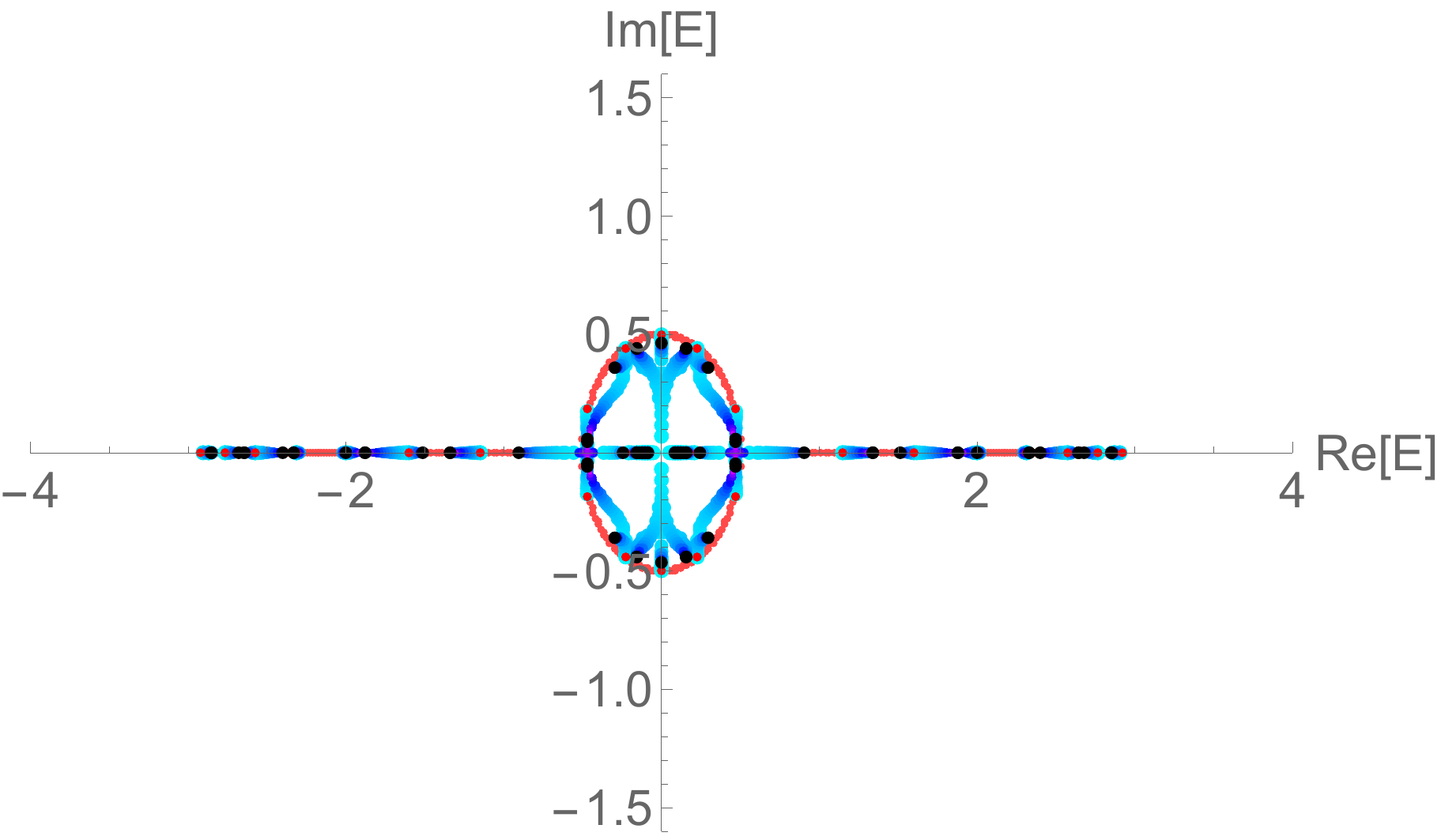}}
			\subfloat[]{\includegraphics[width=.32\linewidth]{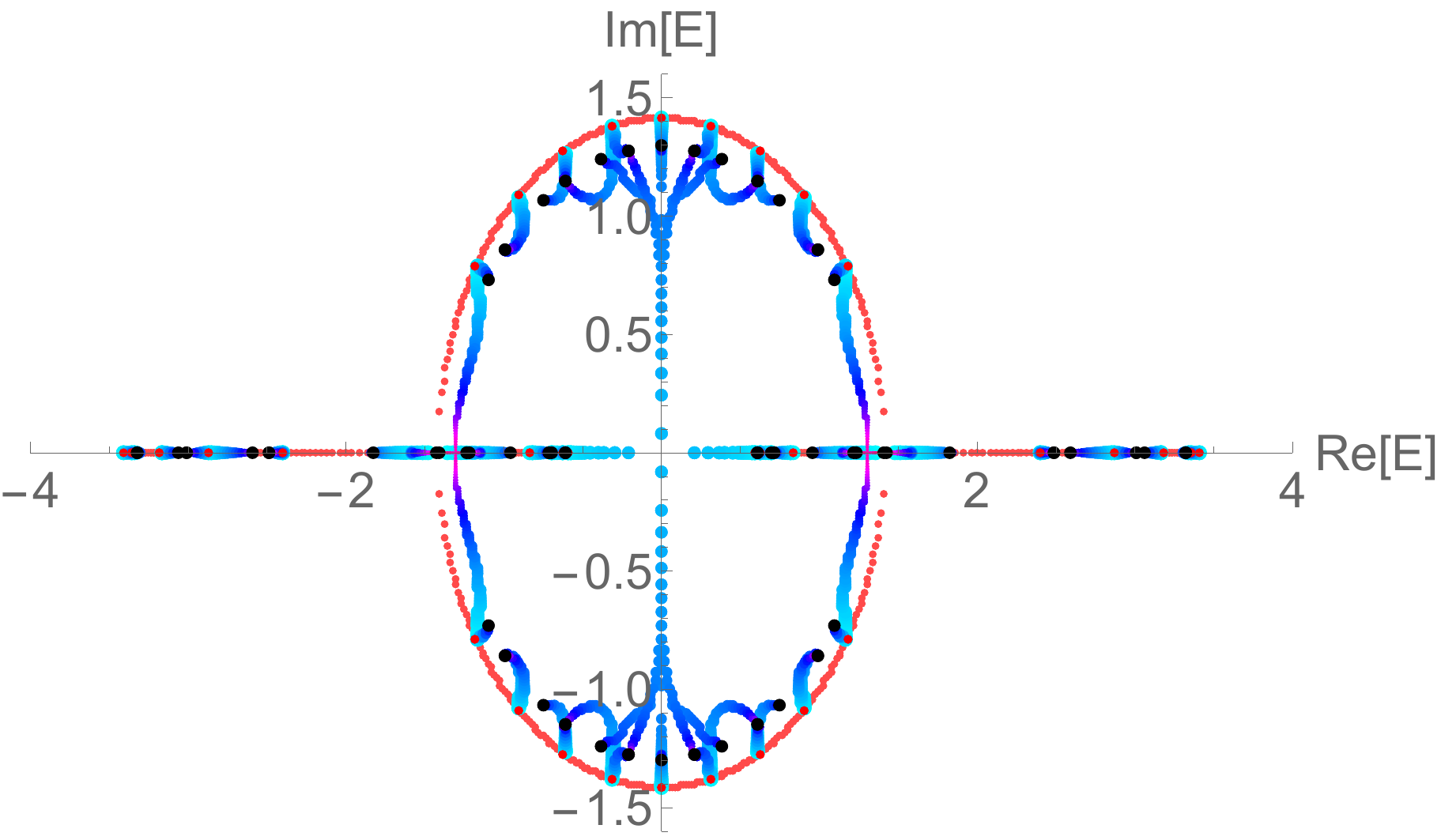}}
			\caption{The PBC-OBC spectral interpolation of the model $H_\text{in}^1$ (Eq.~\ref{Hin1}) for parameters $A=B=\sqrt{2},1.5$ and $2$ for (a-c) respectively. Red/black dots represent PBC/OBC eigenenergies. The blue-purple curves represent the evolution of selected eigenenergies as PBCs are continuously deformed into OBCs by turning off the boundary hoppings. For this model, once a complex ``bubble'' appears in the PBC spectrum, the OBC spectrum ceases to be real too. }
			\label{ABspects}
		\end{figure}

		\section{Error tolerance in determining the parameter region of real spectra}
		In general, the real spectra of our models were determined numerically. In most cases, it is clear whether the collapse onto the real energy line occurs, and the parameter region does not depend on the numerical error tolerance $\epsilon$ for $\text{Max}\text{Im}(E)$. However, in some cases that are potentially afflicted with the critical skin effect~\cite{li2020critical,liu2020helical,rafi2021critical}, there is strong sensitivity to system size, and different $\epsilon$ also gives rise to different parameter regions for real spectra. Shown below are a few illustrative cases.

		\begin{figure}
			\includegraphics[width=.19\linewidth]{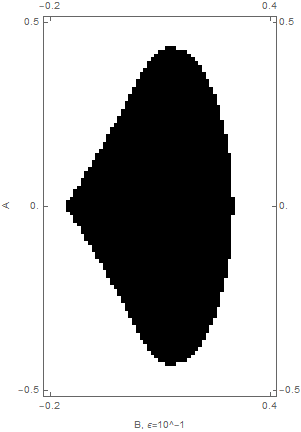}
			\includegraphics[width=.19\linewidth]{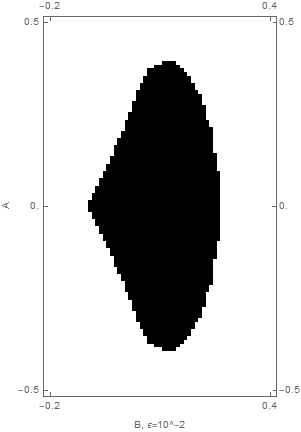}
			\includegraphics[width=.19\linewidth]{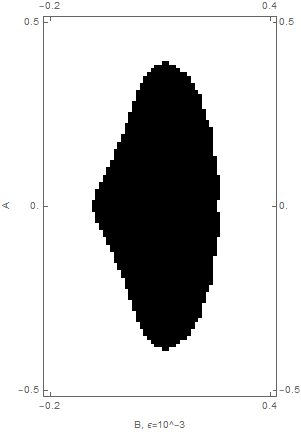}
			\includegraphics[width=.19\linewidth]{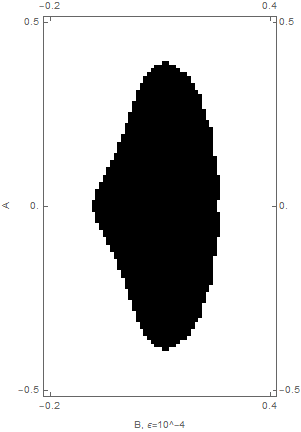}
			\includegraphics[width=.19\linewidth]{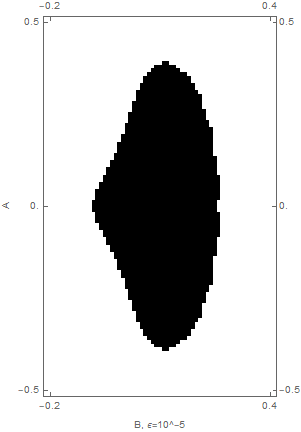}\\
			\includegraphics[width=.19\linewidth]{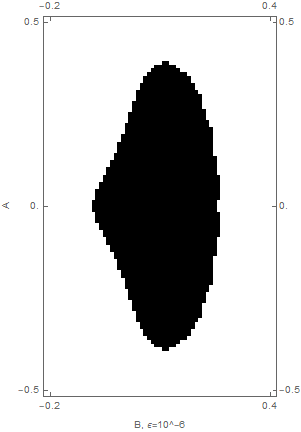}
			\includegraphics[width=.19\linewidth]{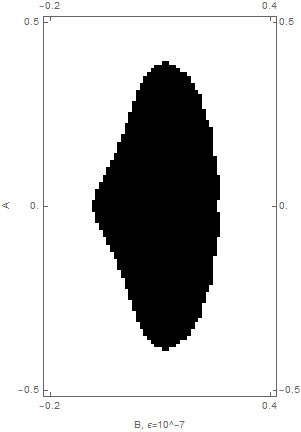}
			\includegraphics[width=.19\linewidth]{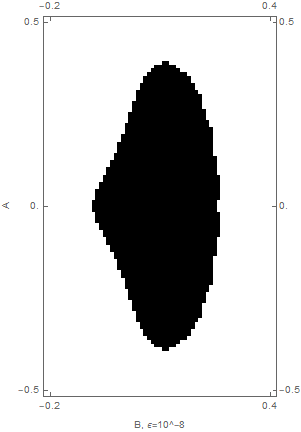}
			\includegraphics[width=.19\linewidth]{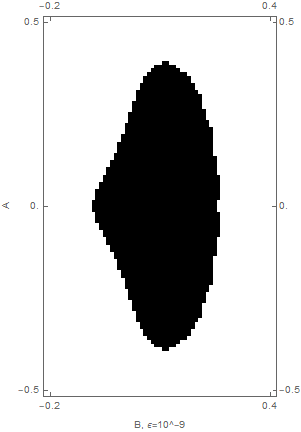}
			\includegraphics[width=.19\linewidth]{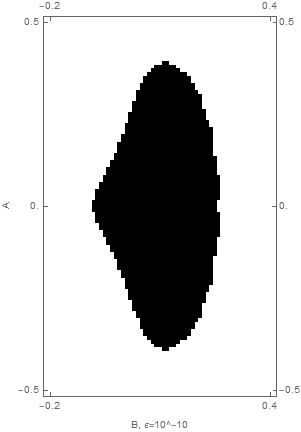}\\
			\includegraphics[width=.19\linewidth]{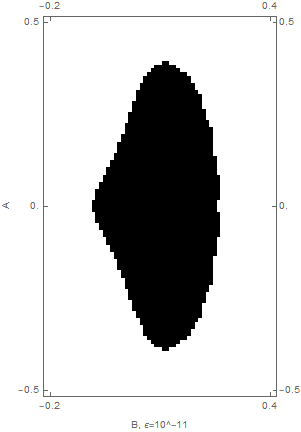}
			\includegraphics[width=.19\linewidth]{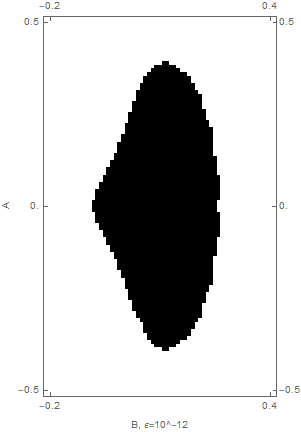}
			\includegraphics[width=.19\linewidth]{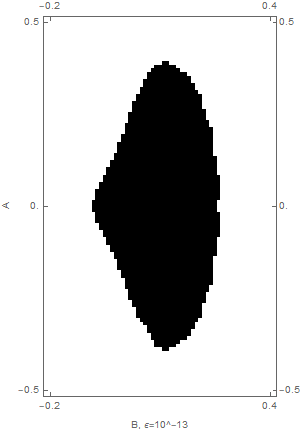}
			\includegraphics[width=.19\linewidth]{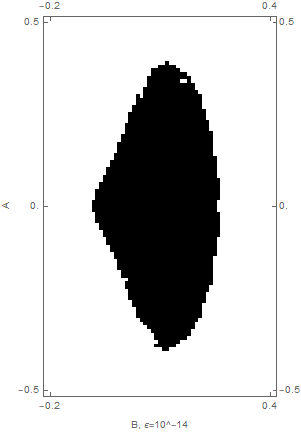}
			\includegraphics[width=.19\linewidth]{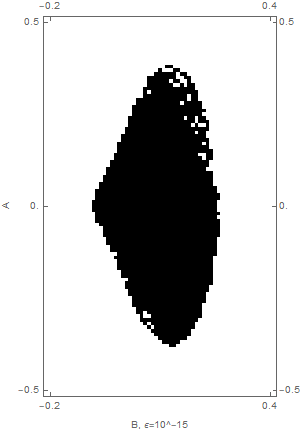}
			\caption{ Dependence of the real spectrum parameter region for the model $H_\text{1-band}^1$ on the threshold $\epsilon$. From top left to bottom right, $\epsilon$ is set $10^{-1}$, $10^{-2}$, $10^{-3}$, ... ,$10^{-15}$ respectively. $\epsilon=10^{-3}$ to $\epsilon=10^{-9}$ give exactly same parameter region, which only exhibits some degeneration beyond $\epsilon=10^{-14}$. This shows a clear independence on the threshold $\epsilon$. 
			}
		\end{figure}

		\begin{figure}
			\includegraphics[width=.19\linewidth]{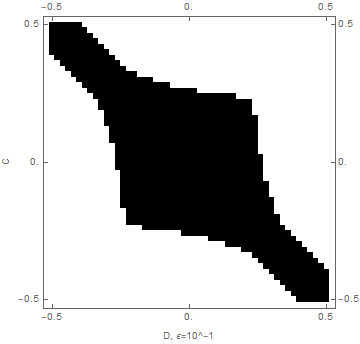}
			\includegraphics[width=.19\linewidth]{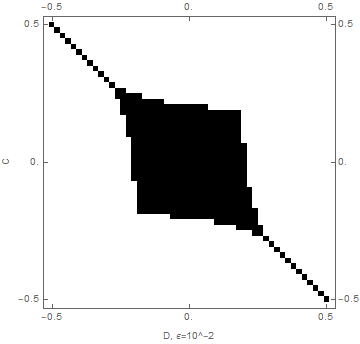}
			\includegraphics[width=.19\linewidth]{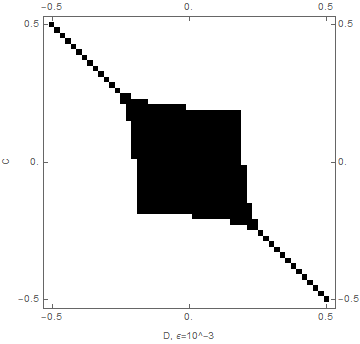}
			\includegraphics[width=.19\linewidth]{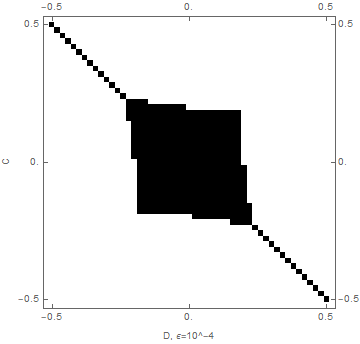}
			\includegraphics[width=.19\linewidth]{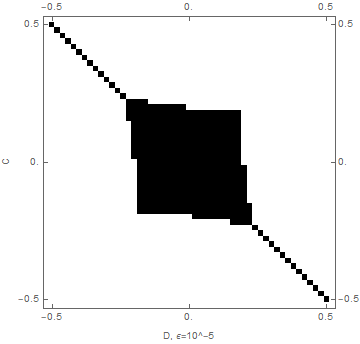}\\
			\includegraphics[width=.19\linewidth]{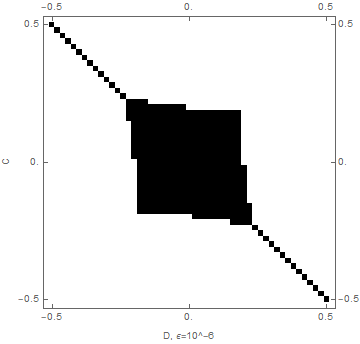}
			\includegraphics[width=.19\linewidth]{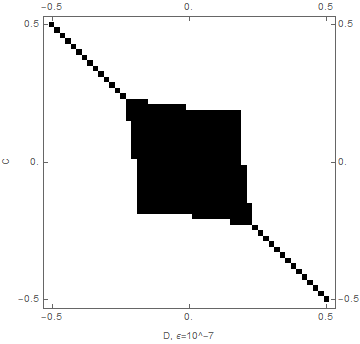}
			\includegraphics[width=.19\linewidth]{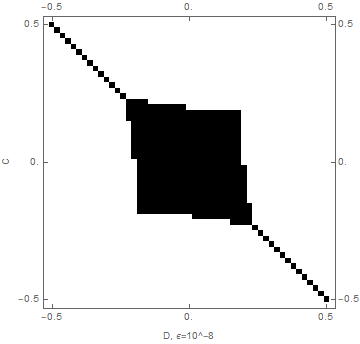}
			\includegraphics[width=.19\linewidth]{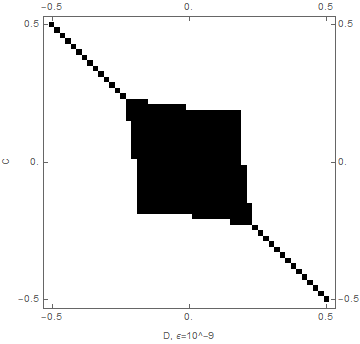}
			\includegraphics[width=.19\linewidth]{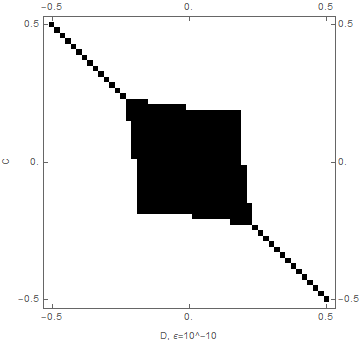}\\
			\includegraphics[width=.19\linewidth]{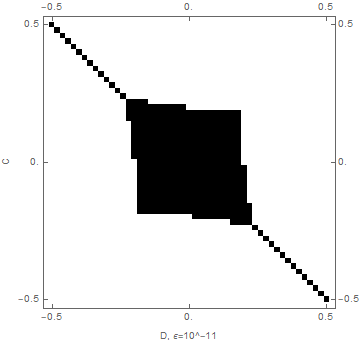}
			\includegraphics[width=.19\linewidth]{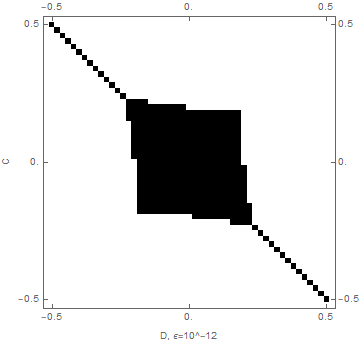}
			\includegraphics[width=.19\linewidth]{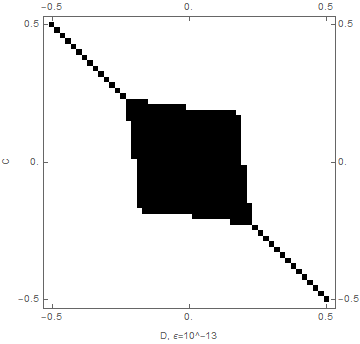}
			\includegraphics[width=.19\linewidth]{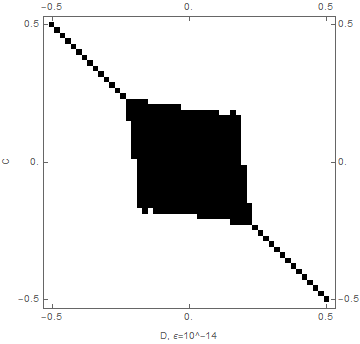}
			\includegraphics[width=.19\linewidth]{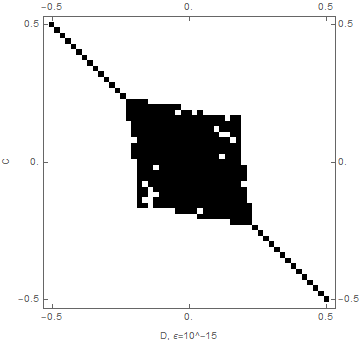}
			\caption{	 Dependence of the real spectrum parameter region for the model $H_\text{1-band}^1$ on the threshold $\epsilon$. From top left to bottom right, $\epsilon$ is set $10^{-1}$, $10^{-2}$, $10^{-3}$, ... ,$10^{-15}$ respectively. $\epsilon=10^{-3}$ to $\epsilon=10^{-12}$ give exactly same parameter region, which only exhibits some degeneration beyond $\epsilon=10^{-14}$. This also shows a clear independence on the threshold $\epsilon$. 
			}
		\end{figure}

	\end{widetext}
	
	
\end{document}